\documentclass[12pt,prd,showpacs,tightenlines,nofootinbib]{revtex4}
\usepackage{bm}
\usepackage{graphics}
\usepackage{rotating}
\usepackage{epsfig}
\begin{document}
\begin{flushright}
{HU-EP-11/32}
\end{flushright}
\title{Exclusive semileptonic and nonleptonic decays of $B$ mesons to 
orbitally excited light mesons}
\author{D. Ebert$^{1}$, R. N. Faustov$^{1,2}$  and V. O. Galkin$^{1,2}$}
\affiliation{
$^1$ Institut f\"ur Physik, Humboldt--Universit\"at zu Berlin,
Newtonstr. 15, D-12489  Berlin, Germany\\
$^2$ Dorodnicyn Computing Centre, Russian Academy of Sciences,
  Vavilov Str. 40, 119991 Moscow, Russia}

\begin{abstract}
The form factors of weak decays of the $B$ meson
to orbitally excited scalar, axial vector and tensor light mesons are
calculated in the framework of the QCD-motivated relativistic quark
model based on the quasipotential approach.  Relativistic effects are
systematically taken into account. The form factors are expressed
trough the overlap integrals of the meson wave functions and their dependence
on the  momentum transfer is selfconsistently determined in the whole
kinematical range. On this basis
semileptonic and two-body nonleptonic $B$ decay rates to orbitally excited
light mesons are calculated. Good agreement of the obtained
predictions with available experimental data is found.

\end{abstract}


\maketitle

\section{Introduction}
\label{sec:int}

Recently significant experimental progress has been achieved in
studying weak decays of $B$ mesons into light mesons \cite{pdg}. Many
new decay modes have been measured including nonleptonic decays involving excited light mesons \cite{expa0,expa1,expb1,expa2}. In
Ref.~\cite{hlsem} we investigated semileptonic $B$ decays to the
ground state $\pi$ and $\rho$ mesons in the framework of the
relativistic quark model based on the quasipotential approach in
QCD. The peculiar feature of the heavy-to-light weak decays is the
very broad kinematical range in which the recoil momentum of the final light
meson is significantly larger than the mass of the light meson except
the small region near the point of zero recoil. Therefore it is very
important to take into account all relevant relativistic effects and
determine the decay form factors without any extrapolations or
additional parameterizations.   Weak decay form factors 
were selfconsistently calculated~\cite{hlsem} in the whole accessible kinematical range. It
was found that both the behaviour of the form factors  on momentum transfer
and differential and total semileptonic decay rates agree well with the
rather precise experimental data.  This allowed us to determine the
Cabibbo-Kobayashi-Maskawa (CKM) matrix element $V_{ub}$.

In this paper we further extend our analysis for the consideration of the
weak semileptonic and two-body nonleptonic $B$ decays to the orbitally
excited light mesons. We calculate the corresponding decay form
factors paying special attention for a consistent account of all
relativistic effects and  determination of the form factor
dependence on the momentum  transfer in the whole accessible
kinematical range. For calculations we use masses and wave 
functions of orbitally excited light mesons which were previously
studied by us in Ref.~\cite{mass}. Table~\ref{tab:nsmm} 
quotes a comparison of
our predictions for the $P$-wave light unflavoured mesons with
experimental data \cite{pdg}. Results for scalar $a_0$, $f_0$, spin
triplet ($^3P_1$) axial vector $a_1$, $f_1$, spin singlet  ($^1P_1$)
axial vector $b_1$, $h_1$ and tensor $a_2$, $f_2$ mesons are
presented. Our model predicts \cite{mass} the light scalar
meson masses heavier than 1~GeV. The scalar mesons below 1~GeV are
well described as the light tetraquarks, composed from the light scalar diquarks
and antidiquarks \cite{ltetr}. From Table~\ref{tab:nsmm} we see that
the calculated mass of the scalar  $q\bar q$ state $1^3P_0$ is significantly
lower than the mass of the  experimentally observed $a_0(1450)$
meson.
Our model predicts that this state is also the tetraquark
composed from the axial vector diquark and antidiquark with the mass
1480~MeV \cite{ltetr}.
Nevertheless, for the purpose of comparison, we here
assume that the lightest scalar 
$q\bar q$ state corresponds to the $a_0(1450)$  meson and test this
assumption by confronting the obtained predictions for the $B$ decays
involving this meson with available experimental data and results of the
different theoretical investigations based on quark models, sum rules, light cone
sum rules and perturbative QCD.  Therefore the study of weak $B$ decays
involving light scalars is an important problem, since it can help to
reveal their real nature.

The calculated weak decay form factors
are used for the evaluation of the semileptonic $B$ decay branching
fractions. Then they are employed for studying the two-body nonleptonic
$B$ decays using the factorization  approach. Such approximation
significantly simplifies calculations, since it 
expresses the matrix elements of the weak 
Hamiltonian responsible for the nonleptonic decays through the product of
the transition matrix elements and meson decay constants. Comparison of the
obtained results with experimental data, which are mostly available for
the nonleptonic $B$ decays involving light axial vector mesons
\cite{expa1,expb1}, can help in testing this approach and discriminate
between different models for form factors.

 \begin{table}
\caption{Predicted \cite{mass} and measured masses of the $P$-wave  light ($q=u,d$) unflavored mesons
   (in MeV).} 
   \label{tab:nsmm}
\begin{ruledtabular}
\begin{tabular}{ccccccc}
&&Theory
&\multicolumn{4}{l}{\underline{\hspace{3.6cm}Experiment \cite{pdg}\hspace{3.6cm}}\hspace*{-2.9cm}}\\
$n^{2S+1}L_J$&$J^{PC}$&$q\bar q$ & $I=1$&mass& $I=0$&mass\\
\hline
$1^3P_0$& $0^{++}$&1176& $a_0$&1474(19)&$f_0$&1200-1500 \\
$1^3P_1$& $1^{++}$&1254& $a_1$&1230(40)&$f_1$&1281.8(6)
\\
$1^3P_2$& $2^{++}$&1317& $a_2$&1318.3(6)&$f_2$&1275.1(12)
\\
$1^1P_1$& $1^{+-}$&1258& $b_1$&1229.5(3.2)&$h_1$&1170(20)
\end{tabular}
\end{ruledtabular}
\end{table}

\section{Relativistic quark model}  
\label{rqm}

In the quasipotential approach a meson is described as a bound
quark-antiquark state with a wave function satisfying the
quasipotential equation of the Schr\"odinger type 
\begin{equation}
\label{quas}
{\left(\frac{b^2(M)}{2\mu_{R}}-\frac{{\bf
p}^2}{2\mu_{R}}\right)\Psi_{M}({\bf p})} =\int\frac{d^3 q}{(2\pi)^3}
 V({\bf p,q};M)\Psi_{M}({\bf q}),
\end{equation}
where the relativistic reduced mass is
\begin{equation}
\mu_{R}=\frac{E_1E_2}{E_1+E_2}=\frac{M^4-(m^2_1-m^2_2)^2}{4M^3},
\end{equation}
and $E_1$, $E_2$ are the center of mass energies on mass shell given by
\begin{equation}
\label{ee}
E_1=\frac{M^2-m_2^2+m_1^2}{2M}, \quad E_2=\frac{M^2-m_1^2+m_2^2}{2M}.
\end{equation}
Here $M=E_1+E_2$ is the meson mass, $m_{1,2}$ are the quark masses,
and ${\bf p}$ is their relative momentum.  
In the center of mass system the relative momentum squared on mass shell 
reads
\begin{equation}
{b^2(M) }
=\frac{[M^2-(m_1+m_2)^2][M^2-(m_1-m_2)^2]}{4M^2}.
\end{equation}

The kernel 
$V({\bf p,q};M)$ in Eq.~(\ref{quas}) is the quasipotential operator of
the quark-antiquark interaction. It is constructed with the help of the
off-mass-shell scattering amplitude, projected onto the positive
energy states. 
Constructing the quasipotential of the quark-antiquark interaction, 
we have assumed that the effective
interaction is the sum of the usual one-gluon exchange term with the mixture
of long-range vector and scalar linear confining potentials, where
the vector confining potential
contains the Pauli interaction. The quasipotential is then defined by
\cite{mass}
  \begin{equation}
\label{qpot}
V({\bf p,q};M)=\bar{u}_1(p)\bar{u}_2(-p){\mathcal V}({\bf p}, {\bf
q};M)u_1(q)u_2(-q),
\end{equation}
with
$${\mathcal V}({\bf p},{\bf q};M)=\frac{4}{3}\alpha_sD_{ \mu\nu}({\bf
k})\gamma_1^{\mu}\gamma_2^{\nu}
+V^V_{\rm conf}({\bf k})\Gamma_1^{\mu}
\Gamma_{2;\mu}+V^S_{\rm conf}({\bf k}),$$
where $\alpha_s$ is the QCD coupling constant, $D_{\mu\nu}$ is the
gluon propagator in the Coulomb gauge
\begin{equation}
D^{00}({\bf k})=-\frac{4\pi}{{\bf k}^2}, \quad D^{ij}({\bf k})=
-\frac{4\pi}{k^2}\left(\delta^{ij}-\frac{k^ik^j}{{\bf k}^2}\right),
\quad D^{0i}=D^{i0}=0,
\end{equation}
and ${\bf k=p-q}$. Here $\gamma_{\mu}$ and $u(p)$ are 
the Dirac matrices and spinors
\begin{equation}
\label{spinor}
u^\lambda_i({p})=\sqrt{\frac{\epsilon_i(p)+m_i}{2\epsilon_i(p)}}
\left(
\begin{array}{c}1\cr {\displaystyle\frac{\bm{\sigma}
      {\bf  p}}{\epsilon_i(p)+m_i}}
\end{array}\right)\chi^\lambda,
\end{equation}
where  $\bm{\sigma}$   and $\chi^\lambda$
are Pauli matrices and spinors and $\epsilon_i(p)=\sqrt{{\bf p}^2+m^2_i}$.
The effective long-range vector vertex is
given by
\begin{equation}
\label{kappa}
\Gamma_{\mu}({\bf k})=\gamma_{\mu}+
\frac{i\kappa}{2m}\sigma_{\mu\nu}k^{\nu},
\end{equation}
where $\kappa$ is the Pauli interaction constant characterizing the
long-range anomalous chromomagnetic moment of quarks. Vector and
scalar confining potentials in the nonrelativistic limit reduce to
\begin{eqnarray}
\label{vlin}
V^V_{\rm conf}(r)&=&(1-\varepsilon)(Ar+B),\nonumber\\ 
V^S_{\rm conf}(r)& =&\varepsilon (Ar+B),
\end{eqnarray}
reproducing 
\begin{equation}
\label{nr}
V_{\rm conf}(r)=V^S_{\rm conf}(r)+V^V_{\rm conf}(r)=Ar+B,
\end{equation}
where $\varepsilon$ is the mixing coefficient. 

The expression for the quasipotential of the heavy quarkonia,
expanded in $v^2/c^2$  can be found in Ref.~\cite{hmass}. The
quasipotential for the heavy quark interaction with a light antiquark
without employing the nonrelativistic ($v/c$)  expansion 
is given in Refs.~\cite{mass,hlm}.  All the parameters of
our model like quark masses, parameters of the linear confining potential
$A$ and $B$, mixing coefficient $\varepsilon$ and anomalous
chromomagnetic quark moment $\kappa$ are fixed from the analysis of
heavy quarkonium masses and radiative
decays  \cite{hmass}. The quark masses
$m_b=4.88$ GeV, $m_c=1.55$ GeV, $m_s=0.5$ GeV, $m_{u,d}=0.33$ GeV and
the parameters of the linear potential $A=0.18$ GeV$^2$ and $B=-0.30$ GeV
have values inherent for quark models.  The value of the mixing
coefficient of vector and scalar confining potentials $\varepsilon=-1$
has been determined from the consideration of the heavy quark expansion
for the semileptonic $B\to D$ decays
\cite{fg} and charmonium radiative decays \cite{hmass}.
Finally, the universal Pauli interaction constant $\kappa=-1$ has been
fixed from the analysis of the fine splitting of heavy quarkonia ${
}^3P_J$- states \cite{hmass} and  the heavy quark expansion for semileptonic
decays of heavy mesons \cite{fg} and baryons \cite{sbar}. Note that the 
long-range  magnetic contribution to the potential in our model
is proportional to $(1+\kappa)$ and thus vanishes for the 
chosen value of $\kappa=-1$ in accordance with the flux tube model.

\section{Matrix elements of the electroweak current} 
\label{mml}

In order to calculate the exclusive semileptonic decay rate of the
$B$ meson, it is necessary to determine the corresponding matrix
element of the  weak current between meson states. 
In the quasipotential approach,  the matrix element of the weak current
$J^W_\mu=\bar q\gamma_\mu(1-\gamma_5)b$, associated with the $b\to q$  transition, between a $B$ meson with mass $M_{B}$ and
momentum $p_{B}$ and a final $P$-wave light meson $F$  with mass $M_F$ and momentum $p_F$ takes the form \cite{f}
\begin{equation}\label{mxet} 
\langle F(p_F) \vert J^W_\mu \vert B(p_{B})\rangle
=\int \frac{d^3p\, d^3q}{(2\pi )^6} \bar \Psi_{F\,{\bf p}_F}({\bf
p})\Gamma _\mu ({\bf p},{\bf q})\Psi_{B\,{\bf p}_{B}}({\bf q}),
\end{equation}
where $\Gamma _\mu ({\bf p},{\bf q})$ is the two-particle vertex function and  
$\Psi_{M\,{\bf p}_M}$ are the
meson ($M=B,F)$ wave functions projected onto the positive energy states of
quarks and boosted to the moving reference frame with momentum ${\bf p}_M$.
\begin{figure}
  \centering
%\vskip 1.5cm
  \includegraphics{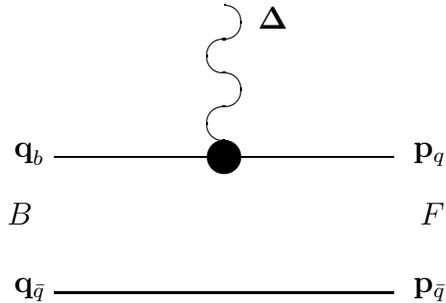}
\caption{Lowest order vertex function $\Gamma^{(1)}_\mu({\bf p},{\bf q})$
contributing to the current matrix element (\ref{mxet}). \label{d1}}
\end{figure}

\begin{figure}
  \centering
%\vskip 1.5cm
  \includegraphics{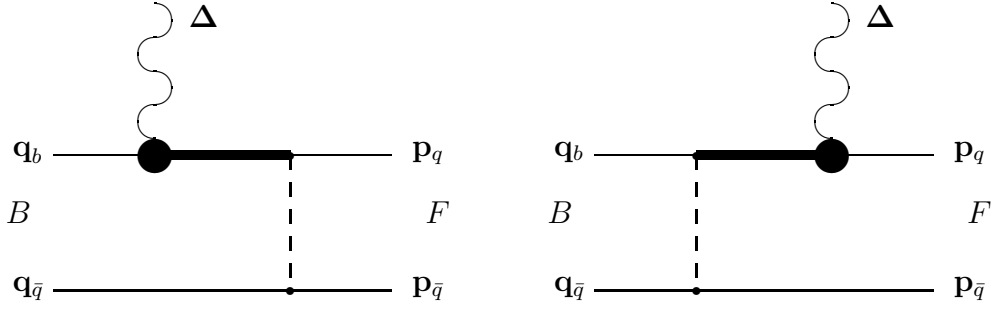}
\caption{ Vertex function $\Gamma^{(2)}_\mu({\bf p},{\bf q})$
taking the quark interaction into account. Dashed lines correspond  
to the effective potential ${\cal V}$ in 
(\ref{qpot}). Bold lines denote the negative-energy part of the quark
propagator. \label{d2}}
\end{figure}

 The contributions to $\Gamma_\mu({\bf p},{\bf q})$ come from Figs.~\ref{d1} and \ref{d2}. The leading order vertex function $\Gamma^{(1)}_\mu({\bf p},{\bf q})$ corresponds to the impulse
approximation, while the vertex function $\Gamma^{(2)}_\mu({\bf p},{\bf q})$ accounts for
contributions of the negative-energy states.
Note that the form of the
relativistic corrections emerging from the vertex function
$\Gamma^{(2)}_\mu({\bf p},{\bf q})$ explicitly depends on the Lorentz structure of the
quark-antiquark interaction. In the leading order of the $v^2/c^2$
expansion for  $B$ and $F$ 
only $\Gamma^{(1)}_\mu({\bf p},{\bf q})$ contributes, while $\Gamma^{(2)}_\mu({\bf p},{\bf q})$  
contributes at the subleading order. 
The vertex functions look like
\begin{equation} \label{gamma1}
\Gamma_\mu^{(1)}({\bf
p},{\bf q})=\bar u_{q}(p_q)\gamma_\mu(1-\gamma^5)u_b(q_b)
(2\pi)^3\delta({\bf p}_{\bar q}-{\bf
q}_{\bar q}),\end{equation}
and
\begin{eqnarray}\label{gamma2} 
\Gamma_\mu^{(2)}({\bf
p},{\bf q})&=&\bar u_{q}(p_q)\bar u_{\bar q}(p_{\bar q}) \Bigl\{\gamma_{1\mu}(1-\gamma_1^5)
\frac{\Lambda_b^{(-)}(
k)}{\epsilon_b(k)+\epsilon_b(p_q)}\gamma_1^0
{\cal V}({\bf p}_{\bar q}-{\bf
q}_{\bar q})\nonumber \\ 
& &+{\cal V}({\bf p}_{\bar q}-{\bf
q}_{\bar q})\frac{\Lambda_{q}^{(-)}(k')}{ \epsilon_{q}(k')+
\epsilon_{q}(q_b)}\gamma_1^0 \gamma_{1\mu}(1-\gamma_1^5)\Bigr\}u_b(q_b)
u_{\bar q}(q_{\bar q}),\end{eqnarray}
where the superscripts ``(1)" and ``(2)" correspond to Figs.~\ref{d1} and
\ref{d2}, the subscripts $q$, $\bar q$, $b$ are the quark indices,  ${\bf k}={\bf p}_q-{\bf\Delta};\
{\bf k}'={\bf q}_b+{\bf\Delta};\ {\bf\Delta}={\bf
p}_F-{\bf p}_{B}$;
$$\Lambda^{(-)}_i(p)=\frac{\epsilon_i(p)-\bigl( m_i\gamma
^0+\gamma^0({\bm{ \gamma}{\bf p}})\bigr)}{ 2\epsilon_i(p)}.$$
Here the quark momenta are expressed through relative momenta ${\bf
  q}$ and $ {\bf p}$ as follows \cite{f} 
\begin{eqnarray*} 
p_{q(\bar q)}&=&\epsilon_{q(\bar q)}(p)\frac{p_F}{M_F}
\pm\sum_{i=1}^3 n^{(i)}(p_F)p^i,\\
q_{b(\bar q)}&=&\epsilon_{b(\bar q)}(q)\frac{p_{B}}{M_{B}} \pm \sum_{i=1}^3 n^{(i)}
(p_{B})q^i,\end{eqnarray*}
and $n^{(i)}$ are three four-vectors given by
$$ n^{(i)\mu}(p)=\left\{ \frac{p^i}{M},\ \delta_{ij}+
\frac{p^ip^j}{M(E+M)}\right\}, \quad E=\sqrt{{\bf p}^2+M^2}.$$

The wave function of a final $P$-wave $F$ meson at rest is given by
\begin{equation}\label{psi}
\Psi_{F}({\bf p})\equiv
\Psi^{J{\cal M}}_{F(^{2S+1}P_J)}({\bf p})={\cal Y}^{J{\cal
    M}}_S\,\psi_{F(^{2S+1}P_J)}({ p}),
\end{equation}
where $J$ and ${\cal M}$ are the total meson angular momentum and its projection,
while $S=0,1$ is the total spin.   
$\psi_{F(^{2S+1}P_J)}({ p})$ is the radial part of the wave function,
which has been determined by the numerical solution of Eq.~(\ref{quas})
in \cite{mass,hlm}.
The spin-angular momentum part ${\cal Y}^{J{\cal M}}_S$ has the following form
\begin{equation}\label{angl}
{\cal Y}^{J{\cal M}}_S=\sum_{\sigma_1\sigma_2}\langle 1\, {\cal M}-\sigma_1-\sigma_2,\  
S\, \sigma_1+\sigma_2 |J\, {\cal M}\rangle\langle \frac12\, \sigma_1,\ 
\frac12\, \sigma_2 |S\, \sigma_1+\sigma_2\rangle Y_{1}^{{\cal M}-\sigma_1-\sigma_2}
\chi_1(\sigma_1)\chi_2(\sigma_2).
\end{equation}
Here $\langle j_1\, m_1,\  j_2\, m_2|J\, {\cal M}\rangle$ are the Clebsch-Gordan 
coefficients, $Y_l^m$ are spherical harmonics, and $\chi(\sigma)$ (where 
$\sigma=\pm 1/2$) are spin wave functions,
$$ \chi\left(1/2\right)={1\choose 0}, \qquad 
\chi\left(-1/2\right)={0\choose 1}. $$

It is important to note that the wave functions entering the weak current
matrix element (\ref{mxet}) are not in the rest frame in general. For example, 
in the $B$ meson rest frame (${\bf p}_{B}=0$), the final  meson
is moving with the recoil momentum ${\bf \Delta}$. The wave function
of the moving  meson $\Psi_{F\,{\bf\Delta}}$ is connected 
with the  wave function in the rest frame 
$\Psi_{F\,{\bf 0}}\equiv \Psi_F$ by the transformation \cite{f}
\begin{equation}
\label{wig}
\Psi_{F\,{\bf\Delta}}({\bf
p})=D_q^{1/2}(R_{L_{\bf\Delta}}^W)D_{\bar q}^{1/2}(R_{L_{
\bf\Delta}}^W)\Psi_{F\,{\bf 0}}({\bf p}),
\end{equation}
where $R^W$ is the Wigner rotation, $L_{\bf\Delta}$ is the Lorentz boost
from the meson rest frame to a moving one, and   
the rotation matrix $D^{1/2}_q(R)$ in spinor representation is given by
\begin{equation}\label{d12}
{1 \ \ \,0\choose 0 \ \ \,1}D^{1/2}_{q}(R^W_{L_{\bf\Delta}})=
S_q^{-1}({\bf p}_{q})S_q({\bf\Delta})S_q({\bf p}),
\end{equation}
where
$$
S_q({\bf p})=\sqrt{\frac{\epsilon_q(p)+m_q}{2m_q}}\left(1+\frac{\bm{\alpha}{\bf p}}
{\epsilon_q(p)+m_q}\right)
$$
is the usual Lorentz transformation matrix of the four-spinor.

\section{Form factors of the semileptonic $B$ decays to the 
  orbitally excited light mesons}
\label{ffr}

The matrix elements of the weak current $J^W_\mu=\bar
b\gamma_\mu(1-\gamma_5)q$ for $B$ decays to orbitally
excited scalar light mesons ($S$) can be parametrized by two invariant
form factors \cite{bcexc}
\begin{eqnarray}
  \label{eq:sff1}
\langle S(p_F)|\bar q \gamma^\mu b|B(p_{B})\rangle
  &=&0,\cr\cr
  \langle S(p_F)|\bar q \gamma^\mu\gamma_5 b|B(p_{B})\rangle
  &=&f_+(q^2)\left(p_{B}^\mu+ p_F^\mu\right)+
  f_-(q^2)\left(p_{B}^\mu- p_F^\mu\right),
\end{eqnarray}
where four-momentum transfer $q=p_{B}-p_F$, 
$M_S$ is the scalar meson mass.

The matrix elements of the weak current for $B$ decays to the spin
triplet ($^3P_1$) axial
vector mesons ($AV$)
can be expressed in terms of four invariant form factors \cite{bcexc}
\begin{eqnarray}
  \label{eq:avff1}
  \langle A(p_F)|\bar q \gamma^\mu b|B(p_{B})\rangle&=&
  (M_{B}+M_A)h_{V_1}(q^2)\epsilon^{*\mu}
  +[h_{V_2}(q^2)p_{B}^\mu+h_{V_3}(q^2)p_F^\mu]\frac{\epsilon^*\cdot q}{M_{B}} ,\qquad\\%\cr
\label{eq:avff2}
\langle A(p_F)|\bar q \gamma^\mu\gamma_5 b|B(p_{B})\rangle&=&
\frac{2ih_A(q^2)}{M_{B}+M_A} \epsilon^{\mu\nu\rho\sigma}\epsilon^*_\nu
  p_{B\rho} p_{F\sigma},  
\end{eqnarray}
where $M_A$ and $\epsilon^\mu$ are the mass and polarization vector of 
the axial vector meson. The matrix elements of the weak current for
$B$ decays to the spin singlet ($^1P_1$) axial vector mesons are
obtained from Eqs.~(\ref{eq:avff1}) by the replacement of the set of
form factors $h_i(q^2)$ by $g_i(q^2)$ ($i=V_1,V_2,V_3,A$).  

The matrix elements of the weak current for $B$ decays to tensor
mesons ($T$)
can be decomposed in four Lorentz-invariant structures \cite{bcexc}
\begin{eqnarray}
  \label{eq:tff1}
  \langle T(p_F)|\bar q \gamma^\mu b|B(p_{B})\rangle&=&
\frac{2it_V(q^2)}{M_{B}+M_T} \epsilon^{\mu\nu\rho\sigma}\epsilon^*_{\nu\alpha}
\frac{p_{B}^\alpha}{M_{B}}  p_{B\rho} p_{F\sigma},\\\cr
\label{eq:tff2}
\langle T(p_F)|\bar q \gamma^\mu\gamma_5 b|B(p_{B})\rangle&=&
(M_{B}+M_T)t_{A_1}(q^2)\epsilon^{*\mu\alpha}\frac{p_{B\alpha}}{M_{B}}\cr%\cr
&&  +[t_{A_2}(q^2)p_{B}^\mu+t_{A_3}(q^2)p_F^\mu]\epsilon^*_{\alpha\beta}
\frac{p_{B}^\alpha p_{B}^\beta}{M_{B}^2} , 
\end{eqnarray}
where $M_T$ and $\epsilon^{\mu\nu}$ are the mass and polarization tensor of 
the tensor meson.

\begin{figure}
  \centering
  \includegraphics[width=7.5cm]{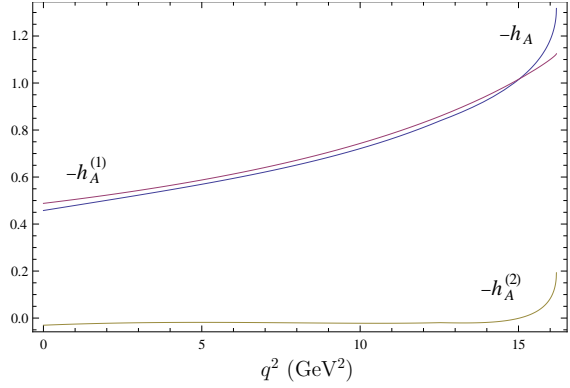}
 \caption{The form factor $h_A(q^2)$ of the $B\to a_1$ weak transition 
   with leading $h_A^{(1)}(q^2)$  and subleading $h_A^{(2)}(q^2)$ contributions.}
  \label{fig:ha}
\end{figure}

The general structure of the current matrix element (\ref{mxet}) is
rather complicated, since it is necessary to integrate both with
respect to $d^3p$ and $d^3q$.
 We calculate exactly the 
contribution of the leading vertex function $\Gamma^{(1)}_\mu({\bf p},{\bf q})$ 
given by Eq. (\ref{gamma1}) to the transition matrix element of the weak
current (\ref{mxet}) using the $\delta$-function. 
As a result the contribution of $\Gamma^{(1)}_\mu({\bf p},{\bf q})$
to the current matrix element has the usual structure of 
an overlap integral of meson wave functions and
can be calculated exactly in the whole kinematical range.
The calculation of the subleading contribution $\Gamma^{(2)}_\mu({\bf p},{\bf q})$ is significantly
more difficult. The heavy quark is  present in the initial $B$ meson
only. Therefore the expansion in its inverse powers retains the dependence on
the relative momentum in the energy of the final light quark. Such
dependence does not  allow one to
perform one of the integrals in the decay matrix element
(\ref{mxet}) using the quasipotential equation. However the final light
meson has a large  recoil momentum
(${\bf\Delta}\equiv{\bf p}_F-{\bf p}_B$, $|{\bf\Delta}_{\rm
  max}|=(M_{B}^2-M_F^2)/(2M_{B})\sim 2.5$~GeV)  almost in
the whole kinematical range except the small region near  
$q^2=q^2_{\rm max}$ ($|{\bf\Delta}|=0$).  This also means that the
recoil momentum of the final meson is large with respect to the mean
relative quark momentum $|{\bf p}|$ in the meson  ($\sim 0.5$~GeV).
Thus one can neglect  $|{\bf p}|$ compared to $|{\bf\Delta}|$ in  the
final light quark energy
$\epsilon_{q}(p+\Delta)\equiv\sqrt{m_{q}^2+({\bf 
p}+{\bf\Delta})^2}$, replacing it by  $\epsilon_{q}(\Delta)\equiv
\sqrt{m_{q}^2+{\bf\Delta}^2}$  in expressions for the
$\Gamma^{(2)}_\mu({\bf p},{\bf q})$.  This replacement removes the relative
momentum dependence in the energy of the light quark and thus permits
to perform one of the integrations in the $\Gamma^{(2)}_\mu({\bf p},{\bf q})$
contribution using the quasipotential equation. This contribution is relatively
small, since it is proportional to the binding energy in the meson. To
demonstrate this observation, we show in Fig.~\ref{fig:ha} leading
$h_A^{(1)}(q^2)$  and subleading $h_A^{(2)}(q^2)$ terms of the form 
factor $h_A(q^2)$ as an example. Contributions of such terms to other
form factors are similar. Therefore application of heavy quark and
large recoil energy expansions and the extrapolation of the
subleading contribution to the small recoil region  introduces
minor errors. Similar calculations for the weak $B$ decays to
light ground state mesons were made in Ref.~\cite{hlsem}. There it was
shown that such extrapolation introduces uncertainties from the
vicinity of zero
recoil $q^2=q^2_{\rm max}$ of less than 1\%. It is important to emphasize that calculating the form
factors we consistently take into account all relativistic 
contributions including the boosts of the meson wave functions from the
rest reference frame to the moving ones, given
by Eq.~(\ref{wig}). Recently we performed the similar calculation of
the weak form factors for the transitions of $B_c$ mesons to the orbitally
excited mesons \cite{bcexc}. Since in this calculation the spectator
charmed quark was treated without the $1/m_c$ expansion and the final
active quark was considered to be light in the
framework of the approach described above, we can use for the present
calculation the expressions
for the decay form factors given in Appendix of Ref.~\cite{bcexc} with
obvious replacements. In
the limits of infinitely heavy quark mass and large recoil energy of the
final meson, these form 
factors satisfy all heavy quark symmetry relations
\cite{clopr,ffhm}.     

 For numerical evaluations of the form factors we use the
quasipotential wave functions of the 
$B$ meson and orbitally excited light mesons
obtained in  \cite{mass,hlm}. Our results for the masses of these
mesons are given in Table~\ref{tab:nsmm}. With the exception of the
$a_0$ meson they are in good agreement
with available experimental data \cite{pdg}.

\begin{figure}
  \centering
  \includegraphics[width=7.5cm]{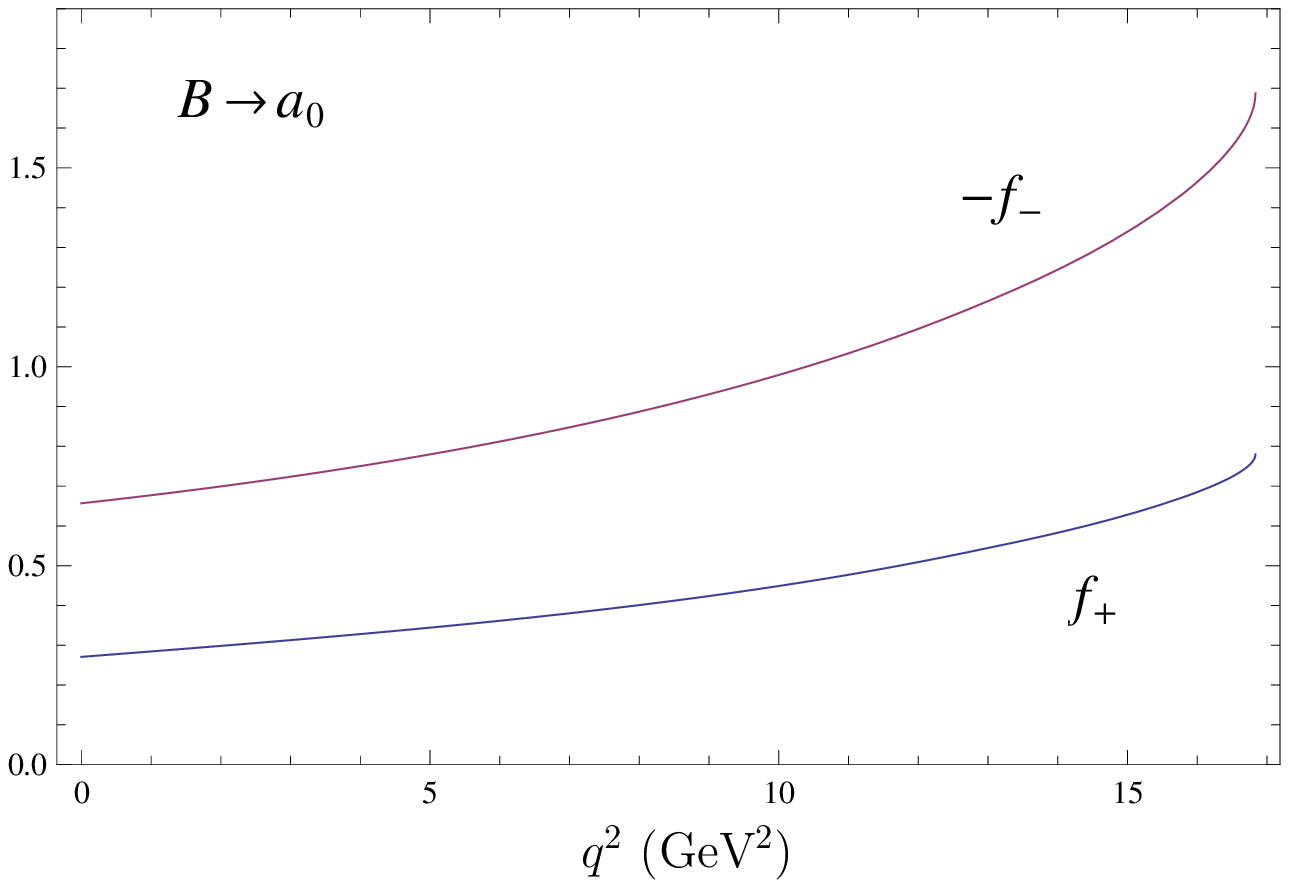} \ \ \ \
\  
\includegraphics[width=7.5cm]{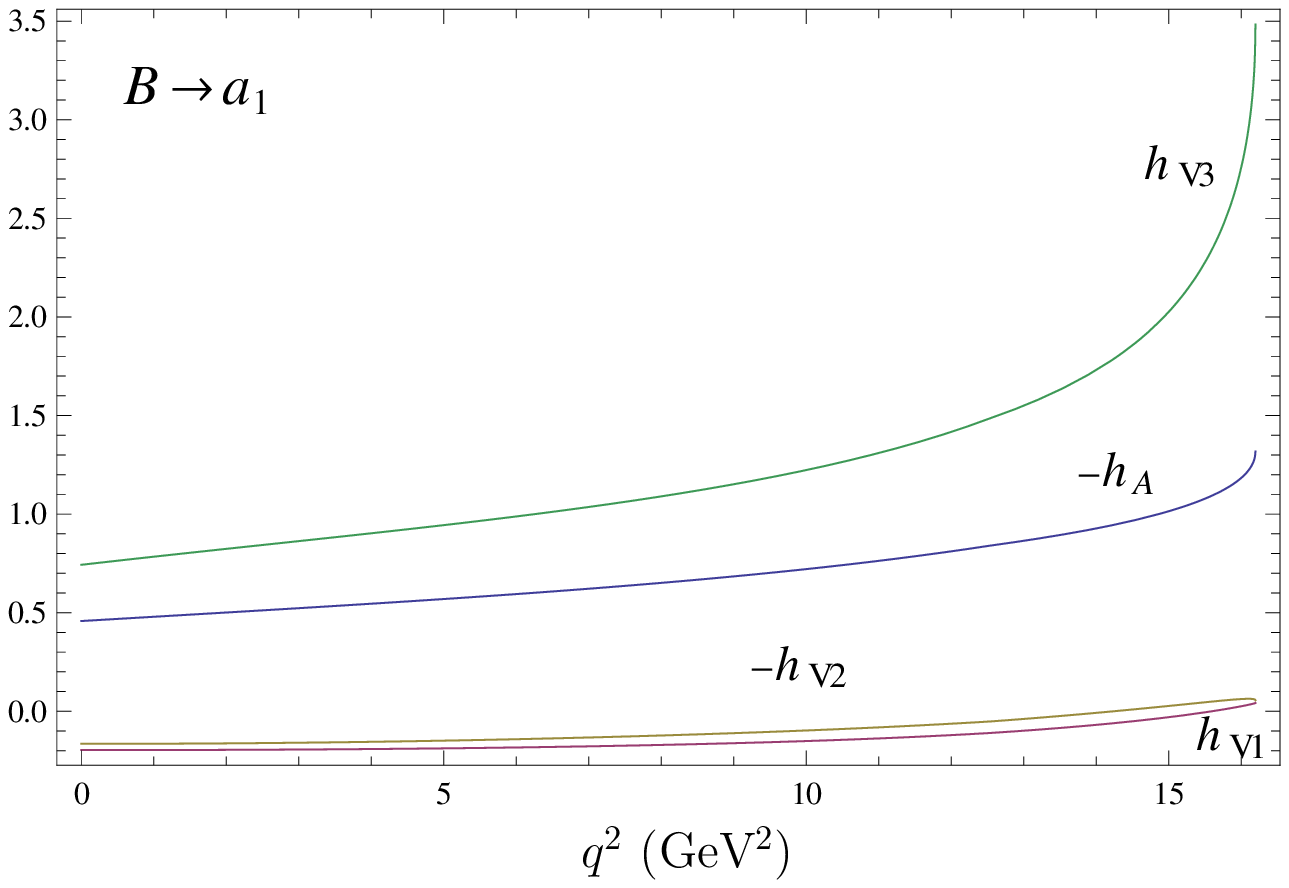}

\vspace*{0.5cm}

  \includegraphics[width=7.5cm]{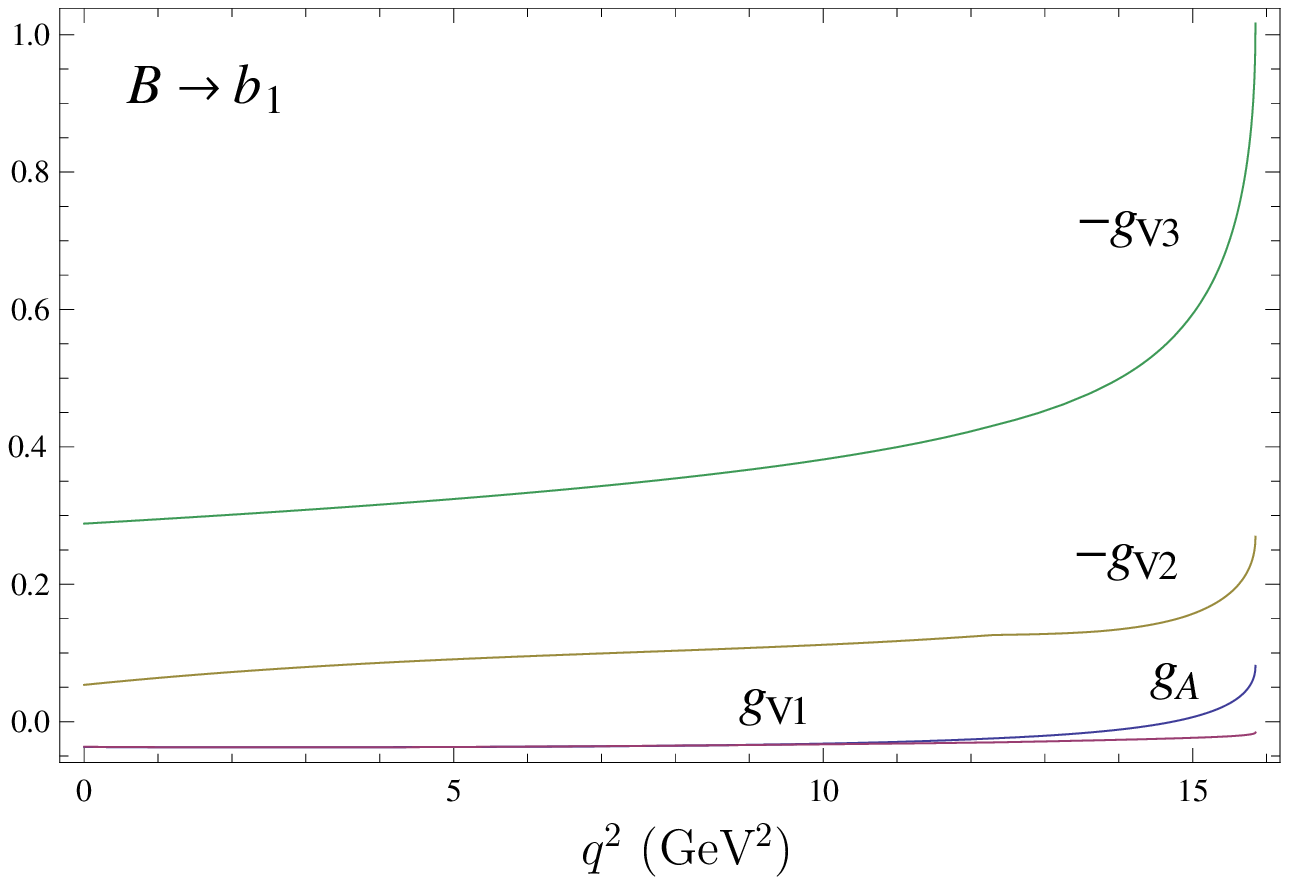} \ \ \ \
\  
\includegraphics[width=7.5cm]{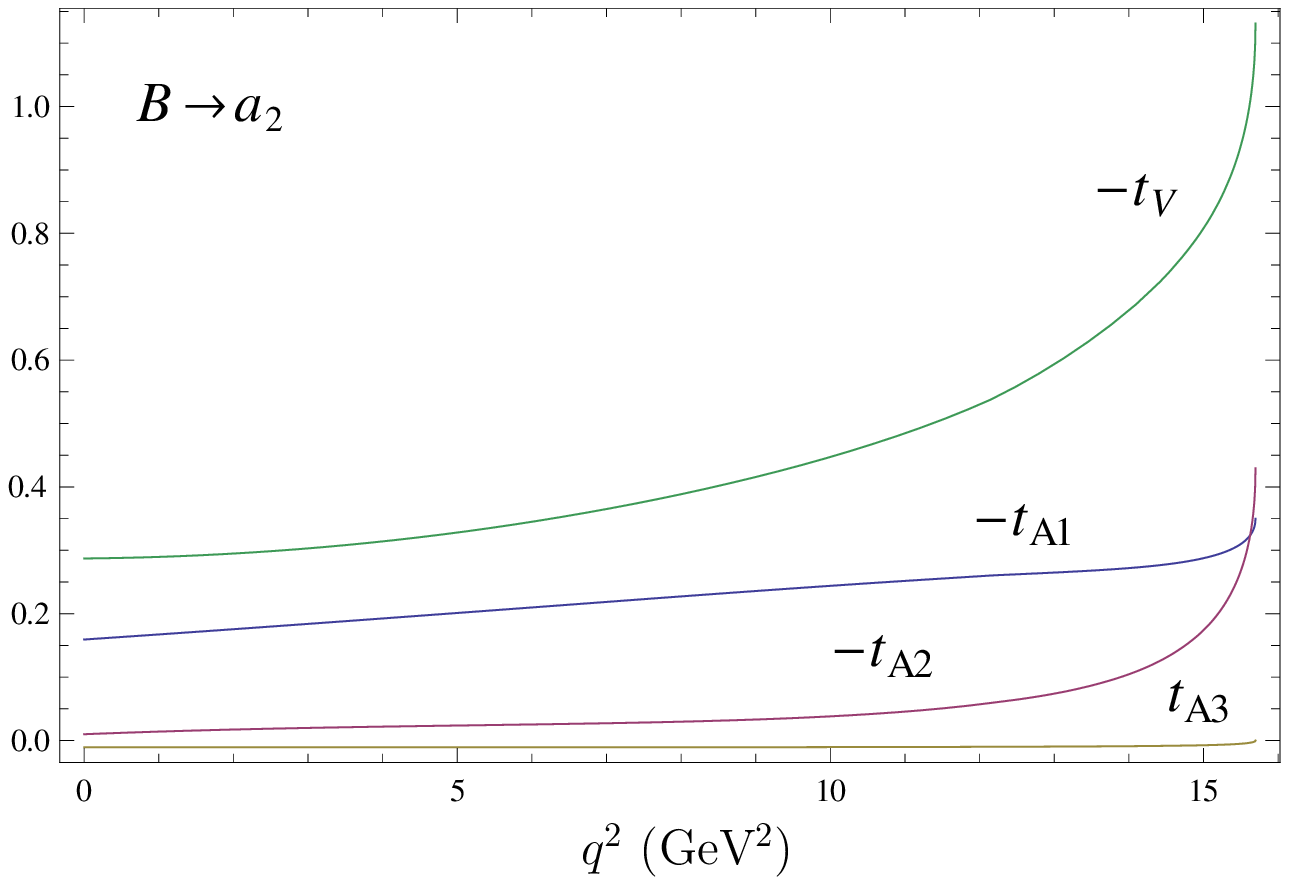}

  \caption{Form factors of the $B$ decays to the $P$--wave light mesons.}
  \label{fig:ffbc}
\end{figure}

In Fig.~\ref{fig:ffbc} we plot form factors of the weak $B$ transitions
to the isovector $P$-wave light mesons (scalar $a_0$, axial vector $a_1$ and
$b_1$, tensor $a_2$). The calculated values of these form factors at
$q^2=0$, several intermediate points  and $q^2=q^2_{\rm
  max}\equiv(M_B-M_F)^2$ are given in Table~\ref{ffm}. 
Since we do not distinguish between isovector and isoscalar mesons, the
form factors for the corresponding weak decays involving isoscalar mesons
coincide with the isovector ones up to the factor $1/\sqrt{2}$, which comes 
from the flavour function of the light neutral meson. As it was argued above, the main
source of the uncertainties of our form factor calculations originates from the
subleading terms (\ref{gamma2}).  We can conservatively estimate the error
 arising from the application of the
heavy quark expansion and the extrapolation of the
subleading contributions to the small recoil region to be less than 1\%
in the interval $q^2=0\div14$~GeV$^2$ and to be less than 4\% for the interval
$q^2=14\ {\rm GeV^2}\div q^2_{\rm max}$. 

These form factors can be approximated with good accuracy by the
following expressions:

\noindent(a) $F(q^2)=
f_+(q^2),f_-(q^2),h_A(q^2),h_{V_3}(q^2),g_{V_1}(q^2),g_{V_2}(q^2),g_{V_3}(q^2),
t_{V}(q^2),t_{A_1}(q^2), t_{A_2}(q^2),$ $t_{A_3}(q^2)$ 
\begin{equation}
  \label{fitfv}
  F(q^2)=\frac{F(0)}{\displaystyle
    \left(1+\sigma_1 
      \frac{q^2}{M_{B}^2}+ \sigma_2\frac{q^4}{M_{B}^4}+\sigma_3 
      \frac{q^6}{M_{B}^6}+ \sigma_4\frac{q^8}{M_{B}^8}\right)},
\end{equation}

\noindent(b) $F(q^2)= h_{V_1}(q^2),h_{V_2}(q^2),g_A(q^2)$
\begin{equation}
  \label{fita12}
F(q^2)=2F(0)-\frac{F(0)}{\displaystyle
    \left(1+\sigma_1 
      \frac{q^2}{M_{B}^2}+ \sigma_2\frac{q^4}{M_{B}^4}+\sigma_3 
      \frac{q^6}{M_{B}^6}+ \sigma_4\frac{q^8}{M_{B}^8}\right)},
\end{equation}
where  the values  $F(0)$ are given in Table~\ref{ffm} and the values
$\sigma_i$ ($i=1,2,3,4$) are given in Table~\ref{hlff}.~\footnote{Note that we do not use 
these parameterizations for further calculations.} In the next
sections we apply the obtained form factors for the calculation of
semileptonic and nonleptonic $B$ decays involving orbitally excited
light mesons.

\begin{table}%[bth]
\caption{Calculated values of the form factors  of the $B$ decays to the $P$--wave
  light mesons  at $q^2=0$, several intermediate points and $q^2=q^2_{\rm max}\equiv(M_B-M_F)^2$. }
\label{ffm}
\begin{ruledtabular}
\begin{tabular}{ccccccccccccccc}
&\multicolumn{2}{c}{\underline{\hspace{0.2cm}$B\to a_0$\hspace{0.2cm}}}&\multicolumn{4}{c}{\underline{\hspace{1.5cm}$B\to a_1$\hspace{1.5cm}}}&\multicolumn{4}{c}{\underline{\hspace{1.5cm}$B\to b_1$\hspace{1.5cm}}}&\multicolumn{4}{c}{\underline{\hspace{1.5cm}$B\to a_2$\hspace{1.5cm}}}\\
$q^2$&$f_+$&$f_-$&$h_A$&$h_{V_1}$&$h_{V_2}$&$h_{V_3}$&$g_A$&$g_{V_1}$&$g_{V_2}$&$g_{V_3}$&$t_V$&$t_{A_1}$&$t_{A_2}$&$t_{A_3}$\\
\hline
0&0.27 &$-0.66$& $-0.46$&$-0.20$&0.17&0.74&$-0.04$ &$-0.04$ & $-0.05$&$-0.29$
&$-0.29$&$-0.16$&$-0.01$&$-0.01$\\
2.5&0.31&$-0.71$&$-0.51$&$-0.20$&0.16&0.84&$-0.04$&$-0.04$&$-0.08$&$-0.31$
&$-0.30$&$-0.18$&$-0.02$&$-0.01$\\
5&0.35&$-0.78$&$-0.57$&$-0.19$&0.15&0.94&$-0.04$&$-0.04$&$-0.09$&$-0.32$
&$-0.33$&$-0.20$&$-0.02$&$-0.01$\\
7.5&0.39&$-0.87$&$-0.64$&$-0.17$&0.13&1.06&$-0.04$&$-0.04$&$-0.10$&$-0.35$
&$-0.38$&$-0.22$&$-0.03$&$-0.01$\\
10&0.45&$-0.98$&$-0.72$&$-0.15$&0.10&1.22&$-0.03$&$-0.03$&$-0.11$&$-0.38$
&$-0.45$&$-0.24$&$-0.04$&$-0.01$\\
12.5&0.53&$-1.13$&$-0.84$&$-0.11$&0.05&1.48&$-0.02$&$-0.03$&$-0.13$&$-0.44$
&$-0.56$&$-0.26$&$-0.06$&$-0.01$\\
15&0.63&$-1.35$&$-1.02$&$-0.03$&$-0.02$&2.05&0.01&$-0.02$&$-0.16$&$-0.60$
&$-0.82$&$-0.29$&$-0.18$&$-0.01$\\
$q^2_{\rm max}$&0.78 &$-1.69$& $-1.32$&0.04& $-0.06$&3.48&$0.08$ & $-0.02$ &$-0.27$ &$-1.02$ 
&$-1.13$& $-0.35$&$-0.43$&0\\
\end{tabular}
\end{ruledtabular}
\end{table}

\begin{table}%[bth]
\caption{Fitted parameters of form factor parameterizations
  (\ref{fitfv}) and (\ref{fita12}).}
\label{hlff}
\begin{ruledtabular}
\begin{tabular}{ccccccccccccccc}
&\multicolumn{2}{c}{\underline{\hspace{0.2cm}$B\to a_0$\hspace{0.2cm}}}&\multicolumn{4}{c}{\underline{\hspace{1.5cm}$B\to a_1$\hspace{1.5cm}}}&\multicolumn{4}{c}{\underline{\hspace{1.5cm}$B\to b_1$\hspace{1.5cm}}}&\multicolumn{4}{c}{\underline{\hspace{1.5cm}$B\to a_2$\hspace{1.5cm}}}\\
&$f_+$&$f_-$&$h_A$&$h_{V_1}$&$h_{V_2}$&$h_{V_3}$&$g_A$&$g_{V_1}$&$g_{V_2}$&$g_{V_3}$&$t_V$&$t_{A_1}$&$t_{A_2}$&$t_{A_3}$\\
\hline
$\sigma_1$&$-1.20$ &$-0.63$& $-0.91$&$0.04$&$-0.15$&$-0.53$&$1.20$ &$-0.95$ & $-3.04$&$0.72$
&$0.98$&$-0.89$&$-4.25$&$-1.33$\\
$\sigma_2$&$-0.27$ &$-2.29$& $-2.40$&$-1.59$&$-1.85$&$-7.10$&$-12.5$ &$9.71$ & $2.46$&$-14.3$
&$-14.7$&$-4.04$&$7.39$&$16.9$\\
$\sigma_3$&$2.35$ &$6.03$& $9.36$&$0.36$&$-0.82$&$23.9$&$38.1$ &$-30.9$ & $13.5$&$45.5$
&$36.8$&$17.0$&$-3.10$&$-63.9$\\
$\sigma_4$&$-2.53$ &$-5.34$& $-9.90$&$-0.93$&$2.52$&$-24.2$&$-41.7$ &$37.1$ & $-22.2$&$-46.4$
&$-31.8$&$-17.4$&$-3.69$&$77.1$\\
\end{tabular}
\end{ruledtabular}
\end{table}

\section{Semileptonic $B$ decays to orbitally excited light mesons}
\label{sec:sd}

The differential decay rate for the $B$ meson decay to $P$-wave light mesons reads \cite{iks}
\begin{equation}
  \label{eq:dgamma}
  \frac{d\Gamma(B\to F(S,AV,T)l\bar\nu)}{dq^2}=\frac{G_F^2}{(2\pi)^3}
  |V_{ub}|^2
  \frac{\lambda^{1/2}(q^2-m_l^2)^2}{24M_{B}^3q^2} 
  \Biggl[H H^{\dag}\left(1+\frac{m_l^2}{2q^2}\right)  +\frac{3m_l^2}{2q^2} H_tH^{\dag}_t\Biggr],
\end{equation}
where $G_F$ is the Fermi constant, $V_{ub}$ is the CKM matrix element, $\lambda\equiv
\lambda(M_{B}^2,M_F^2,q^2)=M_{B}^4+M_F^4+q^4-2(M_{B}^2M_F^2+M_F^2q^2+M_{B}^2q^2)$,
$m_l$ is the lepton mass and
\begin{equation}
  \label{eq:hh}
  H H^{\dag}\equiv H_+H^{\dag}_++H_-H^{\dag}_-+H_0H^{\dag}_0.
\end{equation}
The
helicity components $H_\pm$, $H_0$ and $H_t$ of the hadronic tensor are expressed through the
invariant form factors.

(a) $B\to S(^3P_0)$ transition
\begin{eqnarray}
  \label{eq:has}
  H_\pm&=&0,\cr
H_0&=&\frac{\lambda^{1/2}}{\sqrt{q^2}}f_+(q^2),\cr
H_t&=&\frac1{\sqrt{q^2}}[(M_{B}^2-M_S^2)f_+(q^2)+q^2f_-(q^2)].
\end{eqnarray}

(b) $B\to AV(^3P_1)$ transition
\begin{eqnarray}
  \label{eq:haav}
  H_\pm&=&(M_{B}+M_{AV})h_{V_1}(q^2) \pm\frac{\lambda^{1/2}}{M_{B}+M_{AV}}h_A,\cr
H_0&=&\frac{1}{2M_{AV}\sqrt{q^2}} \left\{(M_{B}+M_{AV})(M_{B}^2-M_{AV}^2-q^2)h_{V_1}(q^2)+\frac{\lambda}{2M_{B}}[h_{V_2}(q^2)+h_{V_3}(q^2)]\right\},\cr
H_t&=&\frac{\lambda^{1/2}}{2M_{AV}\sqrt{q^2}}
\Biggl\{(M_{B}+M_{AV})h_{V_1}(q^2)+\frac{M_{B}^2-M_{AV}^2}{2M_{B}}[h_{V_2}(q^2)+h_{V_3}(q^2)]\cr
&&+\frac{q^2}{2M_{B}}[h_{V_2}(q^2)-h_{V_3}(q^2)]\Biggr\}.
\end{eqnarray}

(c) $B\to AV'(^1P_1)$ transition\\
\nopagebreak
\indent $H_i$ are obtained from
expressions (\ref{eq:haav}) by replacement of form factors $h_i(q^2)$
by $g_i(q^2)$.

(d) $B\to T(^3P_2)$ transition
\begin{eqnarray}
  \label{eq:haat}
  H_\pm&=&\frac{\lambda^{1/2}}{2\sqrt{2}M_{B}M_{T}}\left[(M_{B}+M_{T})t_{A_1}(q^2) \pm\frac{\lambda^{1/2}}{M_{B}+M_{T}}t_V\right],\cr
H_0&=&\frac{\lambda^{1/2}}{2\sqrt{6}M_{B}M_{T}^2\sqrt{q^2}} \left\{(M_{B}+M_{T})(M_{B}^2-M_{T}^2-q^2)t_{A_1}(q^2)+\frac{\lambda}{2M_{B}}[t_{A_2}(q^2)+t_{A_3}(q^2)]\right\},\cr
H_t&=&\sqrt{\frac23}\frac{\lambda}{4M_{B}M_{T}^2\sqrt{q^2}}
\Biggl\{(M_{B}+M_{T})t_{A_1}(q^2)+\frac{M_{B}^2-M_{T}^2}{2M_{B}}[t_{A_2}(q^2)+t_{A_3}(q^2)]\cr
&&+\frac{q^2}{2M_{B}}[t_{A_2}(q^2)-t_{A_3}(q^2)]\Biggr\}.
\end{eqnarray}
Here the subscripts $\pm,0,t$ denote transverse, longitudinal and time helicity
components, respectively.

\begin{figure}
  \centering
 \includegraphics[width=8cm]{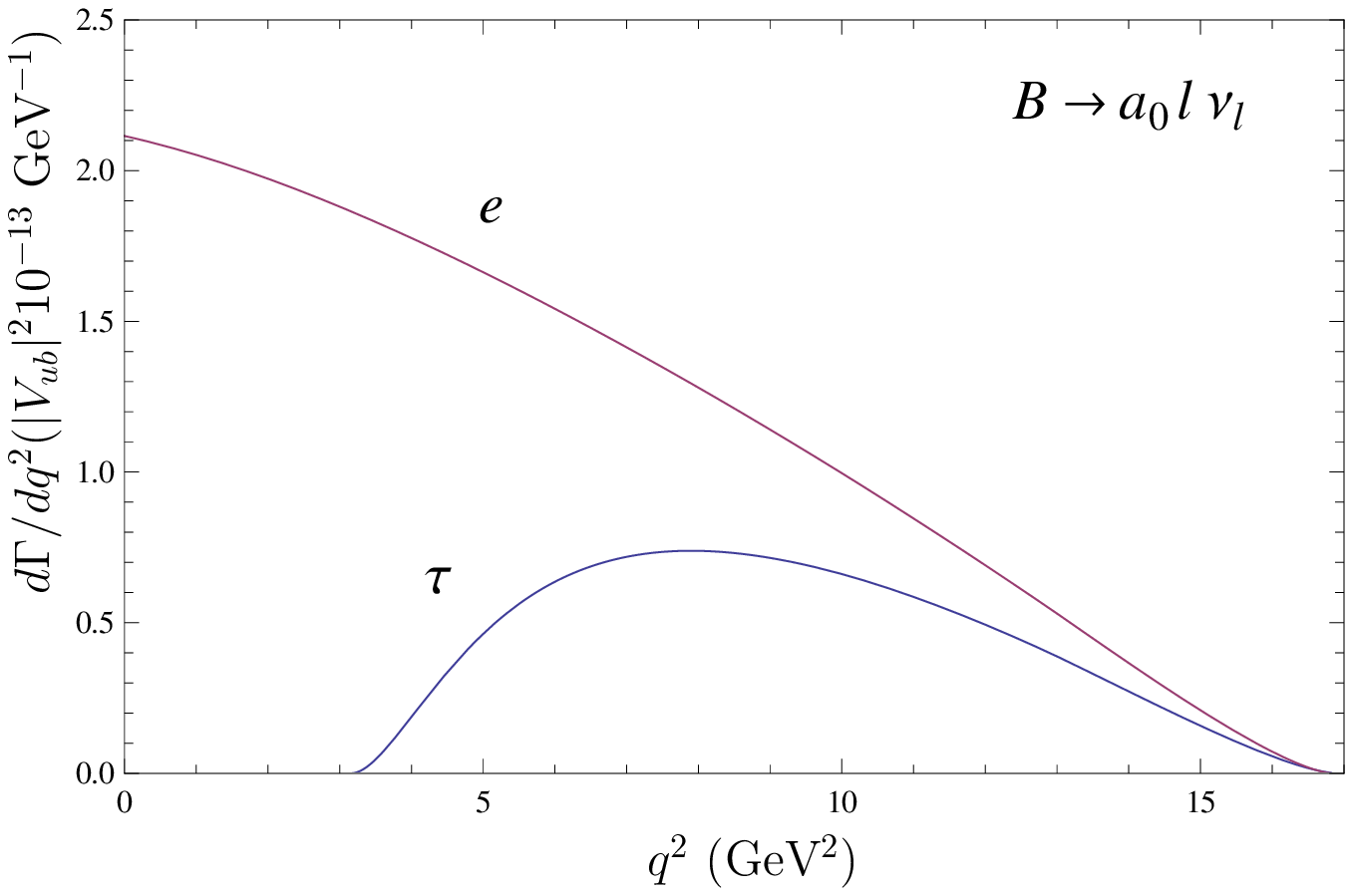}\ \
 \  \includegraphics[width=8cm]{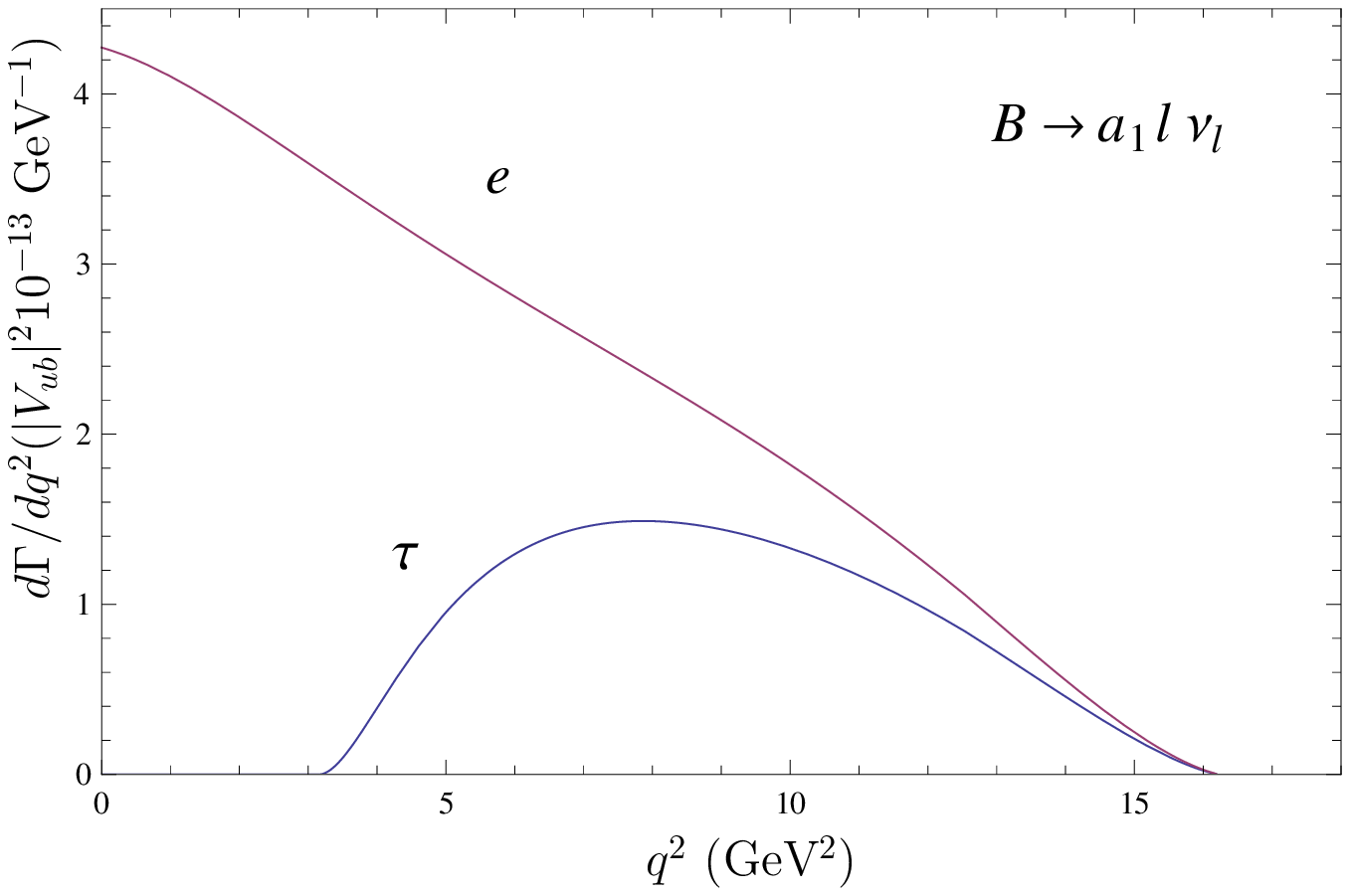}

\vspace*{0.5cm}
 \includegraphics[width=8cm]{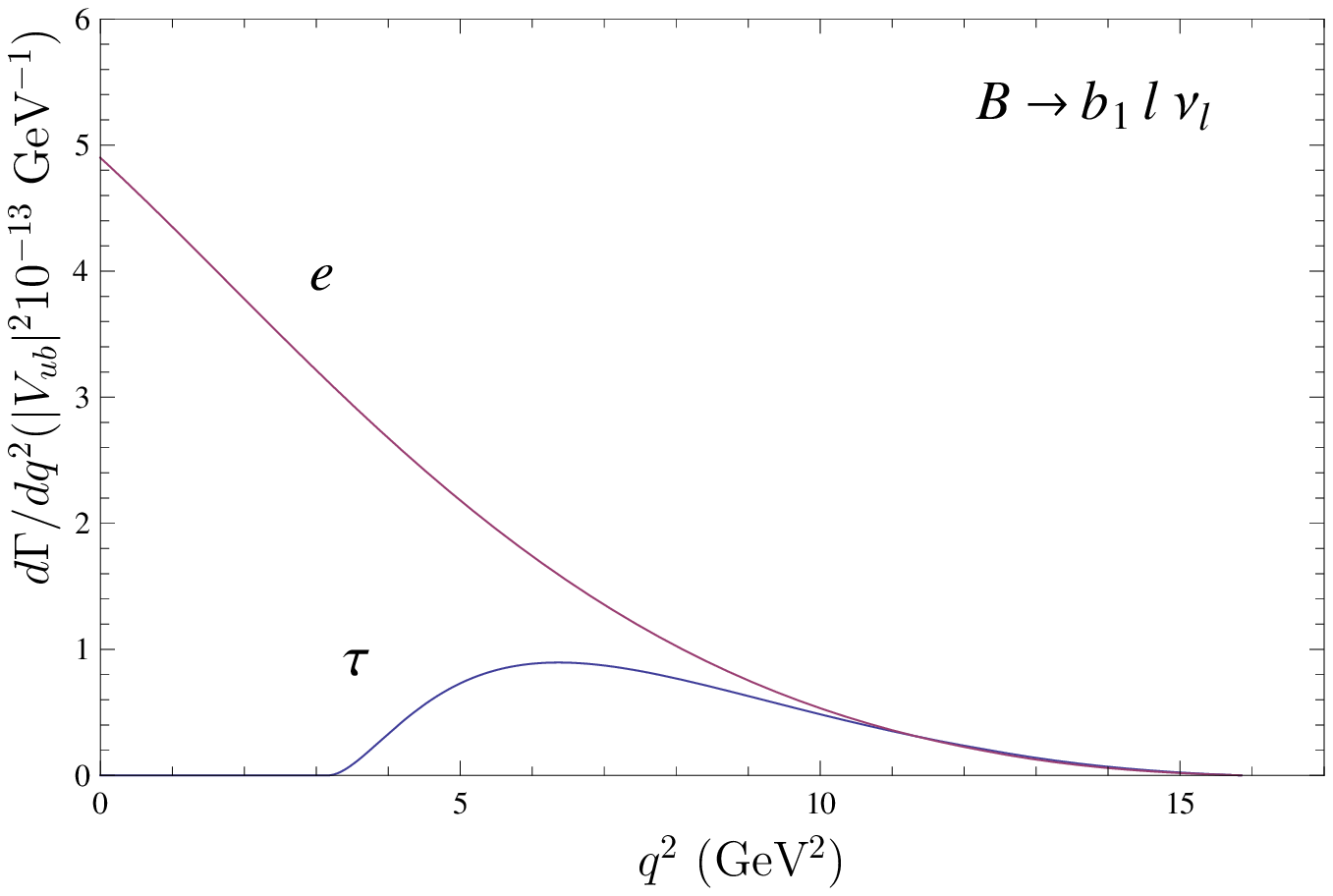}\ \ \  \includegraphics[width=8cm]{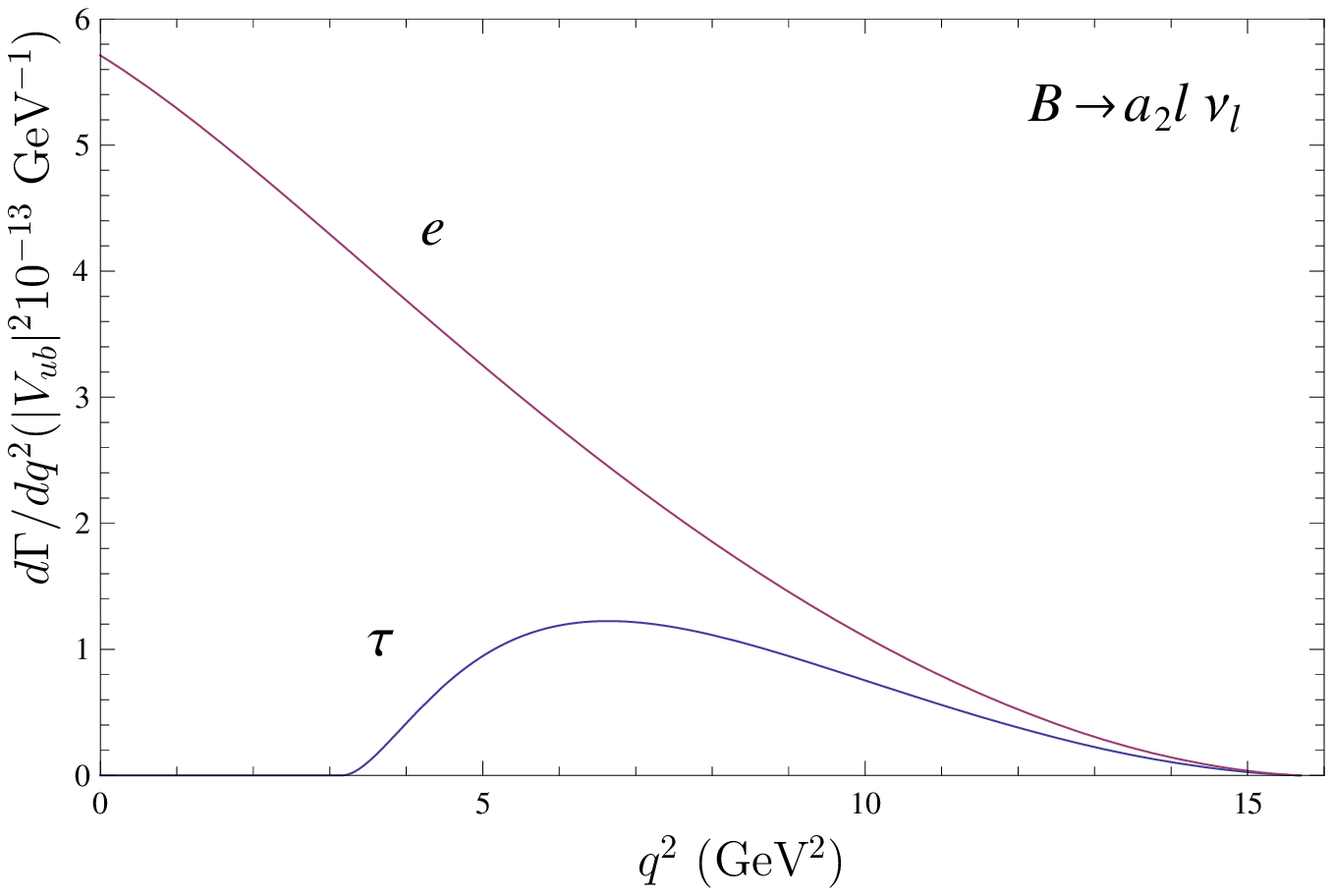}

  \caption{Predictions for the differential decay rates (in $|V_{ub}|^210^{-13}$ GeV$^{-1}$) of the $B$ 
    semileptonic decays to the $P$-wave light mesons. }
  \label{fig:brbc}
\end{figure}

Now we substitute the weak decay form factors calculated in the
previous section in the above expressions for decay
rates. The resulting differential distributions for the $B$ decays to
the $P$-wave light mesons
are plotted in Fig.~\ref{fig:brbc}.  The corresponding total decay
rates are obtained  by integrating
the differential decay rates over $q^2$. For
calculations we use the following value of the CKM matrix element
$|V_{ub}|=0.0038\pm0.00044$ \cite{pdg}. It is necessary to point out that the kinematical
range  accessible in these semileptonic decays is rather
broad. Therefore the knowledge of the $q^2$ dependence of the form
factors is very important for reducing theoretical uncertainties of the decay
rates. Our results for the semileptonic $B$ decay rates to the
$P$-wave light mesons  are given in Table~\ref{dr}.\footnote{We used the
calculated value of the $a_0$ mass from Table~\ref{tab:nsmm}.} 
The errors of the decay rates presented in this table originate from
the form factor uncertainties, discussed in the previous section, and are about 7\%.

\begin{table}%[bth]
\caption{Predictions for the decay rates of the
  semileptonic $B$ decays 
 to the $P$-wave light mesons (in $|V_{ub}|^2$ps$^{-1}$). }
\label{dr}
\begin{ruledtabular}
\begin{tabular}{ccc}
 Decay& $\Gamma(l=e,\mu)$ & $\Gamma(l=\tau)$\\
\hline
$B\to a_0 l\nu$ & 3.18 & 1.00  \\
$B \to a_1l\nu$ & 5.90 & 1.95  \\
$B\to b_1l\nu$ & 4.04 & 0.91  \\
$B\to a_2 l\nu$ & 5.77 & 1.27 \\
\end{tabular}
\end{ruledtabular}
\end{table}

The predictions for the  branching ratios of the
semileptonic $B$ decays to the $P$-wave light meson
states~\footnote{The presented errors of our calculations arise from
the theoretical uncertainties in form factor calculations and
experimental uncertainties which mainly originate from the rather poor
knowledge of the CKM matrix element $V_{ub}$. The latter uncertainty
is dominant.} are
compared with the previous calculations \cite{is,llww,lcsr,a} in
Table~\ref{brb}. The authors of Refs.~\cite{is} use the constituent quark
(ISGW2) model. Perturbative QCD (pQCD) approach is adopted in Ref.~\cite{llww}.
Calculations in 
Ref.~\cite{lcsr} are based on the light-cone QCD sum rules (LCSR), while
Ref.~\cite{a} employs the  QCD sum rules. Our predictions for $B\to
a_{0,1}(b_1)l\nu$ decays are somewhat lower than the central values of pQCD
and LCSR, but they are consistent within rather large errors of the considered 
approaches.  The predicted central values of the branching ratio for the $B\to
a_2l\nu$ decay are close in all calculations except the ISGW2
model, which predicts an order of magnitude lower value. Therefore we find
that essentially different theoretical approaches give the values
for the $B\to a_J(b_1)l\nu$ decay branching ratios of order of
$10^{-4}$ which is the same as for the decays to the ground
state $\pi$ and $\rho$ mesons. It is important to verify these
predictions experimentally.   

\begin{table}%[bth]
\caption{Comparison of theoretical predictions for the branching ratios of the
  semileptonic $B$ decays 
 to the $P$-wave light mesons (in $10^{-5}$). EFG: Ebert, Faustov,
 Galkin, this paper.}
\label{brb}
\begin{ruledtabular}
\begin{tabular}{cccccc}
 Decay& EFG &ISGW2 \cite{is} &pQCD \cite{llww}& LCSR \cite{lcsr}& SR \cite{a}\\
\hline
$\bar B^0\to a_0^+ e\nu$ & $7.0\pm2.8$ & 7.3 & $32.5^{+23.6}_{-13.6}$ & $18^{+9}_{-6}$ & \\
$\bar B^0 \to a_0^+ \tau\nu$ & $2.2\pm0.9$ &  &$13.2^{+9.7}_{-5.7}$ & $6.3^{+3.4}_{-2.5}$ & \\
$\bar B^0 \to a_1^+ e\nu$ & $13.0\pm5.2$ & 19.2 & $29.6^{+17.4}_{-13.9}$ &
$30.2^{+10.3}_{-10.3}$ & 16 \\
$\bar B^0\to a_1^+ \tau\nu$ & $4.3\pm1.7$ &  &$13.4^{+7.8}_{-6.3}$ &  & \\
$\bar B^0\to b_1^+ e\nu$ & $8.9\pm3.6$ & 24.1 & $28.8^{+15.1}_{-12.2}$ & $19.3^{+8.4}_{-6.8}$ & \\
$\bar B^0\to b_1^+ \tau\nu$ & $2.0\pm0.8$ &  &$12.6^{+6.6}_{-5.4}$ &  & \\
$\bar B^0\to a_2^+ e\nu$ & $12.7\pm5.1$ & 1.1 & $11.6^{+8.1}_{-5.7}$ &  16 \\
$\bar B^0\to a_2^+ \tau\nu$ & $2.8\pm1.1$ &  &$4.1^{+2.9}_{-2.0}$ &   6 \\
\end{tabular}
\end{ruledtabular}
\end{table}

\section{Nonleptonic decays}\label{nl}
In the standard model nonleptonic $B$ decays are described by the
effective Hamiltonian, obtained by integrating out the heavy $W$-boson
and top quark. For $\Delta B=1$ transitions $(q=d,s)$ \cite{bbl}
\begin{equation}
\label{heff}
H_{\rm eff}=\frac{G_F}{\sqrt{2}}\Biggl\{V_{cb}V_{cq}^*\left[c_1(\mu)O_1^{c}+
c_2(\mu)O_2^{c}\right] +
V_{ub}V_{uq}^*\left[c_1(\mu)O_1^{u}+
c_2(\mu)O_2^{u}\right]
-V_{tb}V_{tq}^*\sum_{i=3}^{10}c_i(\mu)O_i(\mu)\Biggr\}.%+{\rm h.c.}
\end{equation}

The Wilson coefficients $c_{i}(\mu)$ are evaluated
perturbatively at the $W$ scale and then are evolved down to the
renormalization scale $\mu\approx m_b$ by the renormalization-group
equations. 
The expressions $O_i$ are 
local four-quark operators which are given by
\begin{eqnarray}
\label{o12}
O_1^{q'}&=& (\bar q' b)_{V-A}(\bar q q')_{V-A}\ , 
\cr \cr 
O_2^{q'} &=&(\bar q_i' b_j)_{V-A}(\bar q_j q'_i)_{V-A}\ , \cr\cr 
O_{3(5)}&=& (\bar q b)_{V-A}\sum_{q'}(\bar q' q')_{V-A,(V+A)}\ ,\cr\cr 
O_{4(6)}&=& (\bar q_i b_j)_{V-A}\sum_{q'}(\bar q'_j
q'_i)_{V-A,(V+A)}\ ,\cr\cr 
O_{7(9)}&=&\frac32 (\bar q b)_{V-A}\sum_{q'}e_{q'}(\bar q'
q')_{V+A,(V-A)}\ ,\cr\cr 
O_{8(10)}&=&\frac32 (\bar q_i b_j)_{V-A}\sum_{q'}e_{q'}(\bar q'_j
q'_i)_{V+A,(V-A)}\ ,
\end{eqnarray}
where $e_q$ denotes the quark electric charge
and the following notations are used
$$(\bar qq')_{V\mp A}=\bar q\gamma_\mu(1\mp\gamma_5)q'.$$

The nonleptonic two-body decay amplitude of a
$B$ meson into light mesons can be expressed through the
matrix element of the effective 
weak Hamiltonian $H_{\rm eff}$~\footnote{Since we make all further
  calculation adopting the naive factorization assumption we neglect
  contributions of charming penguins \cite{cfms} which are absent in
  this approximation.} 
\begin{eqnarray}
\label{mel} 
M(B\to M_1M_2)=\langle M_1M_2|H_{\rm eff}|B\rangle&=& \frac{G_F}{\sqrt{2}}
\Biggl\{
V_{ub}V_{uq}^*\left[c_1\langle M_1M_2|O_1^{u}|B\rangle+
c_2\langle M_1M_2|O_2^{u}|B\rangle\right]\cr&& -V_{tb}V_{tq}^*\sum_{i=3}^{10}c_i\langle M_1M_2|O_i(\mu)|B\rangle\Biggr\}.
\end{eqnarray}
The factorization approach, which is extensively used for the calculation
of two-body nonleptonic decays assumes that the
nonleptonic decay amplitude reduces to the product of a meson
transition matrix element
and a decay constant \cite{bsw}. This assumption in general cannot be
exact.  However, it is expected that factorization can hold 
for energetic decays, where both final mesons are light and therefore
possess large recoil momenta \cite{dg}. A justification of this assumption is
usually based on the issue of color 
transparency \cite{jb}. In these decays the final hadrons, which have a
large relative momentum,
are produced in the form of almost point-like color-singlet objects
that do not couple to soft gluons at leading order. Therefore they
undergo only hard interactions with the $B$ meson remnants before they
hadronize.  A more general treatment of factorization is given in
Refs.~\cite{bbns} where it is shown that its naive form follows in the heavy quark
limit at zeroth order in $\alpha_s$ and $\Lambda_{\rm QCD}/m_B$.

Then the decay
amplitude can be approximated by the product of one-particle matrix
elements, e.g., the tree part of the matrix element $(q=d,s)$ is
given by
\begin{eqnarray}
  \label{eq:fact}
  \langle M_1^0M_2^-|c_1O_1+ c_2O_2|B^-\rangle &\approx&
a_1\langle M_1^0|(\bar u b)_{V-A}|B^-\rangle \langle M_2^-|(\bar q
u)_{V-A}|0\rangle\cr\cr  
&&+ a_2\langle M_2^-|(\bar q b)_{V-A}|B^-\rangle \langle M_1^0|(\bar u
u)_{V-A}|0\rangle,
\end{eqnarray}
in which the  Wilson coefficients appear in the following
linear combinations
\begin{eqnarray}
\label{amu} 
a_i&=&c_i+\frac{1}{N_c}c_{i+1}\qquad (i={\rm odd})\cr
a_i&=&c_i+\frac{1}{N_c}c_{i-1} \qquad (i={\rm even})
\end{eqnarray}
and $N_c$ is the number of colors. For calculations we use the
values of the next-to-leading order Wilson coefficients
obtained in Ref.~\cite{bbl} for $\Lambda_{\bar{\rm MS}}^{(5)}=225$~MeV in
the HV ('t Hooft–Veltman) scheme:
$c_1=1.105$, $c_2=-0.228$, $c_3=0.013$,
$c_4=-0.029$, $c_5=0.009$, $c_6=-0.033$,
$c_7/\alpha=0.005$, $c_8/\alpha=0.060$,
$c_9/\alpha=-1.283$, $c_{10}/\alpha=0.266$, where $\alpha$ is
the fine structure constant.

The matrix elements of the weak current $ \langle M|(\bar q
b)_{V-A}|B\rangle$ between $B$ and light meson states
are expressed through the decay form factors (see, e.g.,
(\ref{eq:sff1})-(\ref{eq:tff2})). The matrix element  $ \langle M|(\bar q_1
q_2)_{V-A}|0\rangle$ between  vacuum and a final pseudoscalar ($P$), vector ($V$),
scalar ($S$) and axial vector ($AV$) meson is parametrized by the decay
constants $f_{P,V,S,AV}$
\begin{eqnarray}
&&\langle P|\bar q_1 \gamma^\mu\gamma_5 q_2|0\rangle=if_Pp^\mu_P, \cr\cr 
&&\langle V|\bar q_1\gamma_\mu q_2|0\rangle=\epsilon_\mu M_Vf_V, \cr\cr 
&&\langle S|\bar q_1 \gamma^\mu q_2|0\rangle=f_Sp^\mu_P, \cr\cr 
&&\langle AV|\bar q_1\gamma_\mu\gamma_5 q_2|0\rangle=\epsilon_\mu M_{AV}f_{AV},
\end{eqnarray}
while the corresponding matrix element for the tensor ($T$) meson vanishes since
\begin{equation}
  \label{eq:tf}
  \langle T|\bar q_1\gamma_\mu q_2|0\rangle\propto \epsilon_{\mu\nu}p^\nu=0 ,
\end{equation}
due to the properties of the polarization tensor $\epsilon_{\mu\nu}$.
The pseudoscalar $f_P$ and vector $f_V$ decay constants were
calculated within our model in Ref.~\cite{fpconst}. It was shown that
the complete account of relativistic effects is necessary
to get agreement with experiment for decay constants especially of
light mesons. The scalar decay constant is proportional to the
difference of the light quark masses and thus exactly vanishes for the
neutral scalar mesons and also vanishes for the charged ones if
isospin symmetry is assumed, $f_{a_0^\pm}\approx f_{a_0^0}=f_{f_0}=0$.
The axial vector $b_1$ ($^1P_1$) meson cannot be produced
from the vacuum by
the axial vector current due to the $G$-parity conservation and,
therefore, $f_{b_1}=0$.
We use the following values of the decay constants: $f_\pi=0.131$~GeV,
$f_\rho=0.220$~GeV, $f_K=0.160$~GeV, $f_{K^*}=0.220$~GeV, $f_{\phi}=0.240$~GeV and $f_{a_1}=f_{f_1}=0.238$~GeV. The relevant CKM
matrix elements \cite{pdg} are $|V_{ud}|=0.975$,  $|V_{us}|=0.225$,
$|V_{ub}|=|0.0019-i\, 0.0033|=0.0038$, 
$|V_{tb}|=0.999$, $|V_{td}|=|0.0080-i\,0.0032|= 0.0086$, $|V_{ts}|=0.0403$.

The matrix elements of the weak current between the $B$ meson and
the final light meson  entering the factorized nonleptonic decay
amplitude (\ref{eq:fact}) are parametrized by the set of decay form
factors. Using the
form factors obtained in Sec.~\ref{ffr} and our previous results for
$B$ decays to the ground-state $\pi$ and $\rho$ mesons \cite{hlsem},
we get predictions for the  branching ratios of
the nonleptonic $B$ decay to orbitally excited light mesons
and display them in Tables~\ref{nldra0}-\ref{nldra2} in 
comparison with other calculations
\cite{c,ccy,dlmr,z,cmv,cy,lnp,wll,klo,mq,cy2,sdv}\footnote{Only central
  values are given for all theoretical predictions which have rather
  large error bars.}  and
available experimental data \cite{pdg,expa0,expa1,expb1,expa2}. 
We can roughly estimate the error of our calculations within adopted
naive factorization approach to be about 40\%. It originates both from
theoretical uncertainties in the form factor and effective Wilson
coefficient calculations  and experimental uncertainties in the values
of CKM matrix elements (which are dominant), decay constants and meson masses.

The difference between charge combinations of the same final mesons in
Tables~\ref{nldra0}-\ref{nldra2}  is sometimes enormous. It originates
mainly from the different set of diagrams for the nonleptonic decay
process involving charged and neutral light mesons. For example,
decays $\bar B^0 \to a_{0,2}^+K^-$ and $B^-\to a_{0,2}^0 K^-$ are tree
dominated, while decays $\bar B^0 \to a_{0,2}^0K^0$ and $B^-\to a_{0,2}^- K^0$
are penguin dominated.  In the case of the decays involving the axial vector
$a_1$ meson the situation is more complicated since its decay constant is
not equal to zero and thus additional diagrams, where $a_1$ is
produced by the weak current from the vacuum, contribute to the nonleptonic decay amplitudes. It is also necessary to take into account additional factors arising form the composition of neutral light unflavored mesons [e.g., $a_i^0=1/\sqrt{2}(d\bar d -u\bar u),$ $i=0,1,2$].

\begin{table}
\caption{The  branching ratios of the two-body nonleptonic $B$ decays involving the
  scalar $1^3P_0$ light mesons (in $10^{-6}$ ). }
\label{nldra0}
\begin{ruledtabular}
\begin{tabular}{lcccccc}
Decay& EFG&\cite{c}&  \cite{ccy}& \cite{dlmr}& \cite{z}  & Exp.~\cite{pdg,expa0}\\
\hline
$\bar B^0\to a^+_0\pi^-$ & 3.6& 20  & 3.1& 8 & & $<2.3/{\cal B}(a_0\to\eta\pi)$\\
$\bar B^0\to a^0_0\pi^0$ & 0.03& & 0.7 & \\
$B^-\to a^0_0\pi^-$ & 2.0& & 2.5 &4 \\
$B^-\to a^-_0\pi^0$ & 0.07& & 1.1 &0.01\\
$\bar B^0\to a^+_0K^-$ & 0.29& & 0.3& 1& &$<3.1/{\cal B}(a_0\to\eta\pi)$\\
$\bar B^0\to a^0_0K^0$ & 0.02& & 0.1 &  \\
$B^-\to a^-_0K^0$ & 0.04& & 0.1 \\
$B^-\to a^0_0K^-$ & 0.15& & 0.2 &0.5  \\
$\bar B^0\to a^+_0\rho^-$ & 10.3&38 & 13.3  & \\
$\bar B^0\to a^0_0\rho^0$ & 0.05& & 3.2  & \\
$B^-\to a^0_0\rho^-$ & 5.5& & 25.4 &\\
$B^-\to a^-_0\rho^0$ & 0.10& & 4.5 &\\
$\bar B^0\to a^+_0K^{*-}$ & 2.0& & 5.3 &&28  \\
$\bar B^0\to a^0_0K^{*0}$ & 0.35& & 2.7 && 14 \\
$B^-\to a^0_0K^{*-}$ & 1.1& & 2.6 &&7.0 \\
$B^-\to a^-_0K^{*0}$ & 0.8& & 7.8 &&30 \\
$\bar B^0\to a^+_0a_1^-$ & 11.4 & \\
$\bar B^0\to a^0_0a_1^0$ & 0.05 \\
$B^-\to a^-_0a_1^{0}$ & 0.11 & \\
$B^-\to a^0_0a_1^-$ & 6.1  & \\
\end{tabular}
\end{ruledtabular}
\end{table}

In Table~\ref{nldra0} we present predictions for the branching ratios
of the two-body nonleptonic $B$ decays involving the $P$-wave light
scalar $q\bar q$ meson $a_0$. We compare our results with predictions
of LCSR \cite{c},       
QCD factorization  with the form factors evaluated in the
light-front quark model \cite{ccy}, naive factorization 
with form factors obtained using QCD sum rules \cite{dlmr}, and pQCD \cite{z} approach. We see that LCSR \cite{c}
and pQCD \cite{z} give the branching fractions which are almost an
order of magnitude larger than our predictions. This is the consequence of
significantly larger values of the form factors $f^{Ba_0}_+(0)=0.46$
in LCSR \cite{c} and  $f^{Ba_0}_+(0)=0.86$ in pQCD \cite{z} approaches compared to our
result $f^{Ba_0}_+(0)=0.27$. On the other hand, our predictions are
consistent with the ones of \cite{ccy} and \cite{dlmr} which both use
$f^{Ba_0}_+(0)=0.26$. In Ref. \cite{ccy} it was argued that the
two-body nonleptonic $B$ decay rates involving light scalars are
significantly different in two- and four-quark pictures of these
mesons.  Therefore experimental measurement of these
nonleptonic decay rates can discriminate between different models for
form factors and help to clarify the nature of light scalars. At
present only experimental upper limits \cite{expa0} are available for two $\bar
B^0$ decay modes to the scalar $a_0(1450)$  and charged pion or kaon,
but unfortunately they involve the unmeasured branching ratio ${\cal
  B}(a_0(1450)\to\eta\pi)$.   

\begin{table}
\caption{The  branching ratios of the two-body nonleptonic $B$ decays involving the
  axial vector $1^3P_1$ light mesons (in $10^{-6}$ ). }
\label{nldra1}
\begin{ruledtabular}
\begin{tabular}{lccccccc}
Decay& EFG&  \cite{cmv} & \cite{cy}& \cite{lnp} & pQCD \cite{wll}& \cite{wll} &Exp.~\cite{pdg,expa1}\\
\hline
$\bar B^0\to a^+_1\pi^-$ & 15.7 & 74.3 & 9.1 & 11.8&12.7&10.7& $13.0\pm 4.3$\\
$\bar B^0\to a^-_1\pi^+$ & 21.1 & 36.7 & 23.4 &12.3 &15.7&17.0& $24.2\pm 5.8$\\
$\bar B^0\to a^\pm_1\pi^\pm$ & 36.8 & 111.0 & 32.5& 24.1&28.3&27.7 & $33\pm 5$\\
$\bar B^0\to a^0_1\pi^0$ & 0.34 & 0.27 & 0.9& 1.7&0.12 &5.5& $<1100$\\
$B^-\to a^0_1\pi^-$ & 11.3 & 43.2 & 7.6& 8.8 &6.7&17.2& $20.4\pm 4.7\pm 3.4$\\
$B^-\to a^-_1\pi^0$ & 13.7 & 13.6 & 14.4& 10.6 &8.1&19.0& $13.2\pm 2.7\pm 2.1$\\
$\bar B^0\to a^+_1K^-$ & 13.2 & 72.2 & 18.3 & 41 &20.6&15.8& $16.3\pm 2.9\pm 2.3$\\
$\bar B^0\to a^0_1K^0$ & 6.0 & 42.3 & 6.9 &25 & 8.0&6.3\\
$B^-\to a^-_1K^0$ & 19.8 & 84.1 & 21.6 &52 &25.5&15.5& $33.2\pm 5.0\pm 4.4$\\
$B^-\to a^0_1K^-$ & 14.3 & 43.4 & 13.9 &28 &15.4&10.5 \\
$\bar B^0\to a^+_1\rho^-$ & 20.7 & 4.3 & 23.9 && &&$<61$\\
$\bar B^0\to a^+_1K^{*-}$ & 3.9 & 0.92 & 10.6 & &\\
$B^-\to a^-_1K^{*0}$ & 0.66 & 0.51 & 11.2 & & &&
$1.3^{+1.1+1.1}_{-1.0-2.6}$\\
$\bar B^0\to a^0_1\phi$ & 0.001 & 0.0005 & 0.01& & \\
$\bar B^0\to a^+_1a_1^-$ & 46.1 & 6.4 & 37.4 &&& &$47.3\pm 10.5\pm 6.3$\\
$\bar B^0\to a^0_1a_1^0$ & 0.81 & 0.1 & 0.5 & &\\
$B^-\to a^-_1a_1^{0}$ & 31.5 & 3.6 & 22.4 &&&&$<13000$\\
$\bar B^0\to a^0_1f_1$ & 0.85 & 0.02 & 0.1 & &\\
$B^-\to a^-_1f_1$ & 17.9 & 3.7 & 12.4 &\\
$\bar B^0\to f_1\pi^0$ & 0.56 & 0.47 & 0.26 & \\
$B^-\to f_1\pi^-$ & 11.6 & 34.1 & 5.2 & \\
$\bar B^0\to f_1K^0$ & 2.9 & 34.7 & 14.6 & \\
$B^-\to f_1K^-$ & 4.9 & 31.1 & 14.8 && &&$<2.0$ \\
\end{tabular}
\end{ruledtabular}
\end{table}

\begin{table}
\caption{The  branching ratios of the two-body nonleptonic $B$ decays involving the
  axial vector $1^1P_1$ light mesons (in $10^{-6}$ ). All experimental
values include the unmeasured branching ratios ${\cal B}(b_1\to\omega\pi)$.}
\label{nldrb1}
\begin{ruledtabular}
\begin{tabular}{lccccccc}
Decay& EFG&  \cite{cmv} & \cite{cy}& \cite{lnp}&  pQCD \cite{wll} & \cite{wll}& Exp.~\cite{pdg,expb1}\\
\hline
$\bar B^0\to b^+_1\pi^-$ & 17.7 & 36.2 & 11.2 &0.7&18.7&7.7 \\
$\bar B^0\to b^-_1\pi^+$ & 0 & 0 & 0.3 & $\approx 0$&1.4&0.6\\
$\bar B^0\to b^\pm_1\pi^\pm$ & 17.7 & 36.2 & 11.5&0.7 & 20.2&8.3&$10.9\pm 1.2\pm 0.9$\\
$\bar B^0\to b^0_1\pi^0$ & 0.18 & 0.15 & 1.1&0.01 & 1.5&1.8&$0.4\pm 0.8\pm 0.2$\\
$B^-\to b^0_1\pi^-$ & 9.5 & 18.6 & 9.6 &0.7 &5.1&5.0&$6.7\pm 1.7\pm 1.0$\\
$B^-\to b^-_1\pi^0$ & 0.62 & 0.29 & 0.4 &0.5 &1.0&2.0&$1.8\pm 0.9\pm 0.2$\\
$\bar B^0\to b^+_1K^-$ & 11.6 & 35.7 & 12.1& 2.0 &42.9&8.5& $7.4\pm 1.0\pm 1.0$\\
$\bar B^0\to b^0_1K^0$ & 4.4 & 19.3 & 7.3 & 4.0 &23.3&4.0&$5.1\pm 1.8\pm 0.5$\\
$B^-\to b^-_1K^0$ & 8.3 & 41.5 & 14.0& 3.0 & 55.0&8.6&$9.6\pm 1.7\pm 0.9$\\
$B^-\to b^0_1K^-$ & 8.2 & 18.1 & 6.2 &0.7& 24.9&4.6&$9.1\pm 1.7\pm 1.0$\\
$\bar B^0\to b^+_1\rho^-$ & 22.7 & 1.6 & 32.1 & \\
$\bar B^0\to b^0_1\rho^0$ & 0.20 & 0.002 & 3.2 & &&& $<3.4$\\
$B^-\to b^-_1\rho^0$ & 0.43 & 0.0005 & 0.9 & &&&$<5.2$\\
$B^-\to b^0_1\rho^-$ & 11.2 & 0.86 & 29.1 & &&&$<3.3$\\
$\bar B^0\to b^+_1K^{*-}$ & 4.1 & 0.32 & 12.5 & \\
$\bar B^0\to b^0_1K^{*0}$ & 0.36 & 0.15 & 6.4 &&& &$<8.0$ \\
$B^-\to b^0_1K^{*-}$ & 2.2 & 0.12 & 12.8& &&&$<6.7$\\
$B^-\to b^-_1K^{*0}$ & 0.7 & 0.18 & 7.0& &&&$<5.9$\\
$\bar B^0\to b^0_1\phi$ & 0.001 & 0.0002 & 0.01 & \\
$\bar B^0\to b^+_1b_1^-$ & 0 & 0 & 1.0 & \\
$\bar B^0\to b^0_1b_1^0$ & 0 & 0 & 3.2 & \\
$B^-\to b^-_1b_1^{0}$ & 0 & 0 & 1.4 &\\
$\bar B^0\to h_1\pi^0$ & 0.22 & 0.16 & 0.16 & \\
$B^-\to h_1\pi^-$ & 9.6 & 18.6 & 1.8 & \\
$\bar B^0\to h_1K^0$ & 4.3 & 19.0 & 10.9 & \\
$B^-\to h_1K^-$ & 11.1 & 19.0 & 11.3 & \\
\end{tabular}
\end{ruledtabular}
\end{table}

In Tables~\ref{nldra1} and \ref{nldrb1} we compare theoretical predictions  for
the two-body nonleptonic $B$ decays involving the axial vector $a_1$ and $b_1$
light mesons with available experimental data. Naive factorization
hypothesis and decay form factors calculated within the ISGW2 model are
used in Ref.~\cite{cmv}. In Ref.~\cite{cy} these decays are
investigated in the framework of QCD factorization  with the light cone 
distribution amplitudes evaluated using QCD sum rules.  The
authors of Ref.~\cite{lnp} employ naive factorization and additional
input of a limited number of experimental data~\footnote{Two
  possible values of the $K_1(1270)$ and $K_1(1400)$ mixing angle are
  considered in 
  Ref.~\cite{lnp}. In Tables~\ref{nldra1} and \ref{nldrb1} we present
  results only for its preferred value $58^\circ$.}, while pQCD and
soft collinear effective theory (SCET), with form factors being fitted
parameters, are applied in
Ref.~\cite{wll}.  The two-body
nonleptonic $B$ decays involving axial vector light mesons are the
best studied experimentally among the decays to excited light mesons. Values
or upper limits are available for almost a half of the decays given in
Tables~\ref{nldra1}, \ref{nldrb1}. Notwithstanding rather large
experimental error bars the existing data can already discriminate
between various theoretical approaches. As it is seen from these
tables our results and predictions of Refs.~\cite{cy},
\cite{lnp}, \cite{wll} (pQCD and  SCET) are consistent with each
other (taking into account rather large error bars) for the most decay
branching ratios and agree with the available experimental data. While
the results of Ref.~\cite{cmv} are in most cases significantly
different and seem to be ruled out by experiment.   

\begin{table}
\caption{The  branching ratios of the two-body nonleptonic $B$ decays involving the
  tensor $1^3P_2$ light mesons (in $10^{-6}$ ). }
\label{nldra2}
\begin{ruledtabular}
\begin{tabular}{lcccccc}
Decay& EFG&  \cite{klo}& \cite{mq} &  \cite{cy2} & \cite{sdv} & Exp.~\cite{pdg,expa2}\\
\hline
$\bar B^0\to a^+_2\pi^-$ & 9.8 & 4.9 & 8.19& 5.2 &13.0 & $<300$\\
$\bar B^0\to a^0_2\pi^0$ & 0.009 & 0.0003 & 0.007& 0.24 &0.18 & \\
$B^-\to a^0_2\pi^-$ & 5.2 & 2.6  & 4.38& 3.0& 6.7 &\\
$B^-\to a^-_2\pi^0$ & 0.19 & 0.001 & 0.015& 0.24 & 0.38 &\\
$\bar B^0\to a^+_2K^-$ & 1.6 & 0.58 & 0.73&9.7& 0.95 & \\
$\bar B^0\to a^0_2K^0$ & 0.02 & 0.005 & 0.014&4.2 & \\
$B^-\to a^-_2K^0$ & 0.05 & 0.011 & 0.015& 8.4 &\\
$B^-\to a^0_2K^-$ & 1.0 & 0.31 & 0.39& 4.9&0.51 &$<45$ \\
$\bar B^0\to a^+_2\rho^-$ & 27.1 & 14.7 & 36.2& 11.3& 36.2 & \\
$\bar B^0\to a^0_2\rho^0$ & 0.23 & 0.003 & 0.03& 0.39& 0.5 & \\
$B^-\to a^0_2\rho^-$ & 14.6 & 7.3 & 19.3& 8.4& 19.4 &\\
$B^-\to a^-_2\rho^0$ & 0.51 & 0.007 & 0.071& 0.82 & 1.1 &$<720$\\
$\bar B^0\to a^+_2K^{*-}$ & 5.0 & 3.5 & 7.25&6.1 &1.9 \\
$\bar B^0\to a^0_2K^{*0}$ & 0.9 & 2.1 & 4.0&3.4 & \\
$B^-\to a^-_2K^{*0}$ & 1.8 & 4.5 & 8.6 & 6.1\\
$B^-\to a^0_2K^{*-}$ & 2.7 & 1.9 & 2.8 & 2.9&1.0\\
$\bar B^0\to a^+_2a_1^-$ & 59.0 &  & 42.5 & \\
$\bar B^0\to a^0_2a_1^0$ & 0.32 &  & 0.04 & \\
$B^-\to a^-_2a_1^{0}$ & 0.69 &  & 0.085 &\\
$B^-\to a^0_2a_1^{-}$ & 31.6 &  & 22.7 &\\
$\bar B^0\to f_2\pi^0$ & 0.09 & 0.0003& & 0.15 &0.19 \\
$B^-\to f_2\pi^-$ & 5.1 & 2.8& & 2.7&7.1 &$1.57\pm 0.42\pm 0.16^{+0.53}_{-0.19}$ \\
$\bar B^0\to f_2K^0$ & 0.11 & 0.005& & 3.4 & &$2.7^{+1.9}_{-0.8}\pm0.9$\\
$B^-\to f_2K^-$ & 1.0 & 0.34& & 3.8&0.54 & $1.33\pm 0.30\pm 0.11^{+0.20}_{-0.32}$ \\
\end{tabular}
\end{ruledtabular}
\end{table}

In Table~\ref{nldra2} our predictions for the branching ratios of the
two-body nonleptonic $B$ decays involving tensor $a_2$ and $f_2$ light
mesons are confronted with other theoretical predictions and 
experimental data. References~\cite{klo} and \cite{mq} employ
generalized factorization  complemented by form factors calculated
in the nonrelativistic ISGW model and covariant light-front approach,
respectively. The authors of Ref.~\cite{cy2} apply the  QCD
factorization, while naive factorization and the improved ISGW2 model
are used in Ref.~\cite{sdv}. From this table we see that theoretical
predictions strongly depend on the adopted approach and the model for
form factors. Experimental data are available only for a few considered decay
modes and represent mostly upper limits. The measurements \cite{expa2} were
recently carried out for two charged $B$ decays involving the tensor
$f_2$ meson and 
the charged pion and kaon as well as for one neutral $B$ decay to $f_2$ and
$K^0$. Our model prediction for $B^-\to f_2K^-$ is in
agreement with experiment, while the ones  for $\bar B^0\to f_2K^0$
and for $B^-\to f_2\pi^-$ are lower and 
larger than experimental values, respectively. However experimental errors are still
large in order to make definite conclusions.

\section{Conclusions}
\label{sec:concl}

Weak  form factors of the $B$ mesons decays to the first orbital
excitations of light mesons were calculated in the framework of the
QCD-motivated relativistic quark model. The form factor dependence on
the momentum  transfer was selfconsistently determined in the whole
accessible kinematical range without applying any additional
parameterizations and extrapolations. All relativistic contributions,
including contributions of the intermediate negative-energy states and
transformations of the wave functions to the moving reference frame
were consistently taken into account. This significantly reduces
theoretical uncertainties of the obtained form factors.  

On this basis the branching ratios of the $B$ semileptonic decays to
orbitally excited light mesons were calculated. Our
predictions were compared with other theoretical calculations based on
the ISGW2 quark model \cite{is}, perturbative QCD \cite{llww}, light cone sum
rules \cite{lcsr} and QCD sum rules \cite{a}. It is important to point
out that in most of the previous approaches the weak form
factors were  calculated in some particular kinematical point or
limited kinematical 
range and than were extrapolated to the whole accessible kinematical
range, which is rather broad for such decays. Thus the ISGW2 quark
model allows the calculation of the form factors at $q^2=q^2_{\rm max}$
and then applies the Gaussian parameterization for them,
while light cone sum rules determine form factors in the range near
$q^2=0$ and, therefore, require extrapolation.  It was found that all these
essentially different approaches predict that semileptonic decays to
orbitally excited light mesons have branching ratios of order
$10^{-4}$ which is the same as for the
decays to the ground state $\pi$ and $\rho$ mesons.  

The obtained form factors were used for the evaluation of the
branching ratios of the two-body nonleptonic decays of $B$ mesons
involving orbitally excited light mesons. The
factorization approach was employed to
express the decay matrix elements 
through the products of the weak form factors and decay
constants. Decays involving scalar $a_0(1450)$,\footnote{Under the
 assumption that it is the $1^3P_0$ $q\bar q$ state.} axial vector
$a_1(1260)$, $f_1(1285)$, $b_1(1235)$, $h_1(1170)$ or tensor
$a_2(1320)$, $f_2(1270)$ and light $\pi$, $\rho$, $K$ and $K^*$ mesons
were considered. Obtained predictions were compared with previous
calculations based on naive and generalized factorization with
form factors obtained in different models, QCD factorization, light
cone sum rules 
and perturbative QCD. It was found that the results significantly depend
on the adopted approach for the calculation of the decay matrix
elements and form factors. Our predictions agree well with
the experimental data, which are mostly available for the decays involving
axial vector light mesons, while some of the previous calculations
significantly deviate from experimental values. Future more precise
and comprehensive data, especially on the semileptonic decays, can help to discriminate between various theoretical
approaches and form factor models.       

\acknowledgements
The authors are grateful to M. M\"uller-Preussker  for support  and to
V. Matveev, V. Savrin and M. Wagner for discussions.  Two of us
(R.N.F. and V.O.G.)  acknowledge  the support by the {\it Deutsche
Forschungsgemeinschaft} under contract Eb 139/6-1.


\begin{thebibliography}{99}
\bibitem{pdg}
  K.~Nakamura  [Particle Data Group],
  %``Review of particle physics,''
  J.\ Phys.\ G {\bf 37}, 075021 (2010).
\bibitem{expa0}
  B.~Aubert {\it et al.}  [BABAR Collaboration],
  %``Search for Neutral B-Meson Decays to a0 pi, a0 K, eta rho0, and eta f0,''
  Phys.\ Rev.\  D {\bf 75}, 111102 (2007).
\bibitem{expa1}
  B.~Aubert {\it et al.}  [BABAR Collaboration],
  %``Measurements of CP-Violating Asymmetries in $B^0 \to$ a+-(1) (1260)
  %$\pi^\mp$ decays,''
  Phys.\ Rev.\ Lett.\  {\bf 98}, 181803 (2007);
  %``Evidence for charged B meson decays to a+-(1)(1260) pi0 and a0(1)(1260)
  %pi+-,''
  Phys.\ Rev.\ Lett.\  {\bf 99}, 261801 (2007); %``Observation of B+ ---> a(1)+(1260) K0 and B0 ---> a(1)-(1260) K+,''
  Phys.\ Rev.\ Lett.\  {\bf 100}, 051803 (2008); %``Observation and Polarization Measurement of B0 ---> a(1)(1260)+ a(1)(1260)-
  %Decay,''
  Phys.\ Rev.\  D {\bf 80}, 092007 (2009); P.~del Amo Sanchez {\it et al.}  [The BABAR Collaboration],
  %``Search for B+ meson decay to a1+ K*0,''
  Phys.\ Rev.\  D {\bf 82}, 091101 (2010).

\bibitem{expb1}
  B.~Aubert {\it et al.}  [The BABAR Collaboration],
  %``Observation of B-meson decays to b(1) pi and b(1) K,''
  Phys.\ Rev.\ Lett.\  {\bf 99}, 241803 (2007); %``Observation of B+ ---> b(1)+ K0 and search for B-meson decays to b(1)0 K0
  %and b(1) pi0,''
  Phys.\ Rev.\  D {\bf 78}, 011104 (2008); %``Search for B-meson decays to b(1 rho) and b(1) K*,''
  Phys.\ Rev.\  D {\bf 80}, 051101 (2009).
\bibitem{expa2}
A.~Garmash {\it et al.}  [Belle Collaboration],
  %``Evidence for large direct CP violation in B+- ---> rho(770)0K+- from
  %analysis of the three-body charmless B+- --->K+- pi+- pi-+ decay,''
  Phys.\ Rev.\ Lett.\  {\bf 96}, 251803 (2006);
B.~Aubert {\it et al.}  [BABAR Collaboration],
  %``Evidence for Direct CP Violation from Dalitz-plot analysis of $B^\pm \to
  %K^\pm \pi^\mp \pi^\pm$,''
  Phys.\ Rev.\  D {\bf 78}, 012004 (2008);
  %``Dalitz Plot Analysis of B+- ---> pi+-pi+-pi-+ Decays,''
  Phys.\ Rev.\  D {\bf 79}, 072006 (2009); 
  %``Time-dependent amplitude analysis of B0 --> K0S pi+ pi-,''
  Phys.\ Rev.\  D {\bf 80}, 112001 (2009).
\bibitem{hlsem}
  D.~Ebert, R.~N.~Faustov and V.~O.~Galkin,
  %``Analysis of semileptonic B decays in the relativistic quark model,''
  Phys.\ Rev.\  D {\bf 75}, 074008 (2007).


\bibitem{mass}
  D.~Ebert, R.~N.~Faustov and V.~O.~Galkin,
  %``Mass spectra and Regge trajectories of light mesons in the relativistic
  %quark model,''
  Phys.\ Rev.\  D {\bf 79}, 114029 (2009).
\bibitem{ltetr}
  D.~Ebert, R.~N.~Faustov and V.~O.~Galkin,
  %``Masses of light tetraquarks and scalar mesons in the relativistic quark
  %model,''
  Eur.\ Phys.\ J.\  C {\bf 60}, 273 (2009).
\bibitem{hmass}
  D.~Ebert, R.~N.~Faustov and V.~O.~Galkin,
  %``Properties of heavy quarkonia and B(c) mesons in the relativistic quark
  %model,''
  Phys.\ Rev.\  D {\bf 67}, 014027 (2003).


\bibitem{hlm}
  D.~Ebert, V.~O.~Galkin and R.~N.~Faustov,
  %``Mass spectrum of orbitally and radially excited heavy-light mesons in  the
  %relativistic quark model,''
  Phys.\ Rev.\  D {\bf 57}, 5663 (1998)
  [Erratum-ibid.\  D {\bf 59}, 019902 (1999)]; D.~Ebert, R.~N.~Faustov and V.~O.~Galkin,
  %``Heavy-light meson spectroscopy and Regge trajectories in the relativistic
  %quark model,''
  Eur.\ Phys.\ J.\  C {\bf 66}, 197 (2010).
\bibitem{fg}
  R.~N.~Faustov and V.~O.~Galkin,
  %``Heavy quark 1/m(Q) expansion of meson weak decay form-factors in the
  %relativistic quark model,''
  Z.\ Phys.\  C {\bf 66}, 119 (1995).
\bibitem{sbar}
  D.~Ebert, R.~N.~Faustov and V.~O.~Galkin,
  %``Semileptonic decays of heavy baryons in the relativistic quark model,''
  Phys.\ Rev.\  D {\bf 73}, 094002 (2006).
\bibitem{f} R. N. Faustov, Ann. Phys. {\bf 78}, 176 (1973); Nuovo
Cimento A {\bf 69}, 37 (1970).
\bibitem{bcexc}
  D.~Ebert, R.~N.~Faustov and V.~O.~Galkin,
  %``Semileptonic and nonleptonic decays of B_c mesons to orbitally excited
  %heavy mesons in the relativistic quark model,''
  Phys.\ Rev.\  D {\bf 82}, 034019 (2010).
\bibitem{clopr}
  J.~Charles, A.~Le Yaouanc, L.~Oliver, O.~Pene and J.~C.~Raynal,
  %``Heavy-to-light form factors in the heavy mass to large energy limit of
  %{QCD},''
  Phys.\ Rev.\  D {\bf 60}, 014001 (1999).
\bibitem{ffhm}
  D.~Ebert, R.~N.~Faustov and V.~O.~Galkin,
  %``Form factors of heavy-to-light B decays at large recoil,''
  Phys.\ Rev.\  D {\bf 64}, 094022 (2001).

\bibitem{iks}
  M.~A.~Ivanov, J.~G.~K\"orner and P.~Santorelli,
%``Semileptonic decays of $B_c$ mesons to charmonium states in a
  %relativistic quark model,''
  Phys.\ Rev.\  D {\bf 71}, 094006 (2005)
  [Erratum-ibid.\  D {\bf 75}, 019901 (2007)].

\bibitem{is}
  D.~Scora and N.~Isgur,
  %``Semileptonic meson decays in the quark model: An update,''
  Phys.\ Rev.\  D {\bf 52}, 2783 (1995).
\bibitem{llww}
  R.~H.~Li, C.~D.~Lu, W.~Wang and X.~X.~Wang,
  %``B ---> S Transition Form Factors in the PQCD approach,''
  Phys.\ Rev.\  D {\bf 79}, 014013 (2009); R.~H.~Li, C.~D.~Lu and W.~Wang,
  %``Transition form factors of B decays into p-wave axial-vector mesons in the
  %perturbative QCD approach,''
  Phys.\ Rev.\  D {\bf 79}, 034014 (2009); W.~Wang,
  %``B to tensor meson form factors in the perturbative QCD approach,''
  Phys.\ Rev.\  D {\bf 83}, 014008 (2011).
\bibitem{lcsr}
  Y.~M.~Wang, M.~J.~Aslam and C.~D.~Lu,
  %``Scalar mesons in weak semileptonic decays of B(s),''
  Phys.\ Rev.\  D {\bf 78}, 014006 (2008); K.~C.~Yang,
  %``Form-Factors of B(u,d,s) Decays into P-Wave Axial-Vector Mesons in the
  %Light-Cone Sum Rule Approach,''
  Phys.\ Rev.\  D {\bf 78}, 034018 (2008);
Z.~G.~Wang,
  %``Analysis of the $B \to K^*_2(1430), a_2(1320), f_2(1270)$ form-factors with
  %light-cone QCD sum rules,''
  arXiv:1011.3200 [hep-ph].
\bibitem{a}
  T.~M.~Aliev and M.~Savci,
  %``Semileptonic B ---> a(1) lepton neutrino decay in QCD,''
  Phys.\ Lett.\  B {\bf 456}, 256 (1999).


\bibitem{bbl}
  G.~Buchalla, A.~J.~Buras and M.~E.~Lautenbacher,
  %``Weak decays beyond leading logarithms,''
  Rev.\ Mod.\ Phys.\  {\bf 68}, 1125 (1996).
\bibitem{cfms}
  M.~Ciuchini, E.~Franco, G.~Martinelli and L.~Silvestrini,
  %``Charming penguins in B decays,''
  Nucl.\ Phys.\  B {\bf 501}, 271 (1997).


\bibitem{bsw} M. Bauer, B. Stech, and M. Wirbel, Z. Phys. C {\bf 34},
103 (1987).
\bibitem{dg} M. J. Dugan and B. Grinstein, Phys. Lett. B {\bf 255}, 583
(1991).
\bibitem{jb} J. D. Bjorken, Nucl. Phys. B (Proc. Suppl.) {\bf 11}, 325
(1989). 
\bibitem{bbns} M. Beneke, G. Buchalla, M. Neubert and C. T. Sachrajda,
Phys. Rev. Lett. {\bf 83}, 1914 (1999); Nucl. Phys. B {\bf 591}, 313
(2000). 
% \bibitem{bs} A. J. Buras and L. Silvestrini, Nucl. Phys. B {\bf 569},
%   3 (2000).
% \bibitem{ccty}
%   Y.~H.~Chen, H.~Y.~Cheng, B.~Tseng and K.~C.~Yang,
%   %``Charmless hadronic two-body decays of B/u and B/d mesons,''
%   Phys.\ Rev.\  D {\bf 60}, 094014 (1999).

\bibitem{fpconst}
  D.~Ebert, R.~N.~Faustov and V.~O.~Galkin,
  %``Relativistic treatment of the decay constants of light and heavy  mesons,''
  Phys.\ Lett.\  B {\bf 635}, 93 (2006).








\bibitem{c}
  V.~Chernyak,
  %``Estimates of flavoured scalar production in B decays,''
  Phys.\ Lett.\  B {\bf 509}, 273 (2001).
\bibitem{ccy}
  H.~Y.~Cheng, C.~K.~Chua and K.~C.~Yang,
  %``Charmless hadronic B decays involving scalar mesons: Implications to the
  %nature of light scalar mesons,''
  Phys.\ Rev.\  D {\bf 73}, 014017 (2006); %``Charmless B decays to a scalar meson and a vector meson,''
  Phys.\ Rev.\  D {\bf 77}, 014034 (2008).
\bibitem{dlmr}
  D.~Delepine, J.~L.~Lucio M., J.~A.~Mendoza S. and C.~A.~Ramirez,
  %``A Consistent Scenario for B to PS Decays,''
  Phys.\ Rev.\  D {\bf 78}, 114016 (2008).
\bibitem{z}
  Z.~Q.~Zhang,
  %``Study of scalar meson a0 (1450) from B $\to$ a0 (1450) K* decays,''
  Phys.\ Rev.\  D {\bf 83}, 054001 (2011).



\bibitem{cmv}
  G.~Calderon, J.~H.~Munoz and C.~E.~Vera,
  %``Nonleptonic two-body B-decays including axial-vector mesons in the final
  %state,''
  Phys.\ Rev.\  D {\bf 76}, 094019 (2007).
\bibitem{cy}
  H.~Y.~Cheng and K.~C.~Yang,
  %``Hadronic charmless B decays B ---> AP,''
  Phys.\ Rev.\  D {\bf 76}, 114020 (2007); H.~Y.~Cheng and K.~C.~Yang,
  %``Branching Ratios and Polarization in B ---> VV, VA, AA Decays,''
  Phys.\ Rev.\  D {\bf 78}, 094001 (2008)
  [Erratum-ibid.\  D {\bf 79}, 039903 (2009)].
\bibitem{lnp}
  V.~Laporta, G.~Nardulli and T.~N.~Pham,
  %``Non leptonic B decays to axial-vector mesons and factorization,''
  Phys.\ Rev.\  D {\bf 74}, 054035 (2006)
  [Erratum-ibid.\  D {\bf 76}, 079903 (2007)]
\bibitem{wll}
  W.~Wang, R.~H.~Li and C.~D.~Lu,
  %``What can we learn from $B\to a_1(1260)(b_1(1235))\pi(K)$ decays?,''
  Phys.\ Rev.\  D {\bf 78}, 074009 (2008).




\bibitem{klo}
  C.~S.~Kim, J.~P.~Lee and S.~Oh,
%``Charmless hadronic decays of B mesons to a pseudoscalar and a tensor
  %meson,''
  Eur.\ Phys.\ J.\  C {\bf 22}, 683 (2002);
%``Hadronic B decays to charmless V T final states,''
  Eur.\ Phys.\ J.\  C {\bf 22}, 695 (2002);
  %``Nonleptonic two-body charmless B decays involving a tensor meson in ISGW2
  %model,''
  Phys.\ Rev.\  D {\bf 67}, 014002 (2003).
\bibitem{mq}
  J.~H.~Munoz and N.~Quintero,
  %``Nonleptonic charmless two-body B decays involving tensor mesons in the
  %covariant light-front approach,''
  J.\ Phys.\ G {\bf 36}, 095004 (2009); J.\ Phys.\ G {\bf 36}, 125002 (2009).

\bibitem{cy2}
  H.~Y.~Cheng and K.~C.~Yang,
  %``Charmless Hadronic B Decays into a Tensor Meson,''
  Phys.\ Rev.\  D {\bf 83}, 034001 (2011).
\bibitem{sdv}
  N.~Sharma, R.~Dhir and R.~C.~Verma,
  %``Decays of bottom mesons emitting tensor meson in final state using ISGW II
  %model,''
  Phys.\ Rev.\  D {\bf 83}, 014007 (2011).













\end{thebibliography}
\end{document}